\documentclass[12pt]{article}
\usepackage{amsmath,amssymb,exscale}
\input epsf

\def\gappeq{\mathrel{\rlap {\raise.5ex\hbox{$>$}}
{\lower.5ex\hbox{$\sim$}}}}
\def\lappeq{\mathrel{\rlap{\raise.5ex\hbox{$<$}}
{\lower.5ex\hbox{$\sim$}}}}

\begin{document}
\topmargin -1.0cm
\oddsidemargin -0.8cm
\evensidemargin -0.8cm
\pagestyle{empty}
\begin{flushright}
UNIL-IPT-02-07\\
IC/2002/43\\
hep-th/0206016\\
\today
\end{flushright}
\vspace*{5mm}

\begin{center}

{\Large\bf  A  Formalism to Analyze the Spectrum
of Brane World Scenarios}
\vspace{1.0cm}

{\large Seif Randjbar-Daemi$^a$\footnote{Email: seif@ictp.trieste.it}  and
Mikhail Shaposhnikov$^b$\footnote{Email:
mikhail.shaposhnikov@ipt.unil.ch}}\\

\vspace{.6cm}

{\it {$^{a}$International Center for Theoretical Physics,
Trieste, Italy}}\\
{\it {$^{b}$Institute of Theoretical Physics, University of
Lausanne,\\
CH-1015 Lausanne, Switzerland}}\\

\vspace{.4cm}
\end{center}

\vspace{1cm}
\begin{abstract}
In this paper we develop a formalism to analyze the spectrum of
small perturbations about arbitrary solutions of Einstein,
Yang-Mills and scalar systems. We consider a  general system of
gravitational, gauge and scalar fields in a $D-$dimensional
space-time and give the bilinear action for the fluctuations of
the fields in the system around an arbitrary solution of the
classical field equations. We then consider warped geometries,
popular in brane world scenarios, and use the light cone gauge to
separate the bilinear action into a totally decoupled spin-two,
-one and -zero fluctuations.  We apply our general scheme to
several examples and discuss in particular localization of abelian
and non-abelian gauge fields of the standard model to branes
generated by scalar fields. We show in particular that the
Nielsen-Olsen string solution gives rise to a normalizable
localized spin-1 field in any number of dimensions.
\end{abstract} 
\vfill

\eject
\pagestyle{empty}
\setcounter{page}{1}
\setcounter{footnote}{0}
\pagestyle{plain}


\section{Introduction}
Consider a general system of fields (gravitational, gauge, scalar and
fermion) in higher-dimensional space-time with the action
\begin{eqnarray}
S = \int d^D x \sqrt{-G}\left(\frac{1}{\kappa^2} R \right.& - &
\frac{1}{4} F_{MN} F^{MN} -(D_M\Phi)^\dagger D^M\Phi
-U(\Phi)
\label{action}
\\
  &+&\left. i[\bar \Psi \Gamma^A E_A ^M (\partial_M - \Omega +
ie A_M T)\Psi + \frac{1}{2} g \bar\Psi \Phi  \Psi] \right)~.
\nonumber
\end{eqnarray}
Here $D$ is the total number of dimensions, $R$ is the scalar
curvature,  $A_M$, $\Psi$ and $\Phi$ are the generic (non-abelian)
gauge, fermion and scalar fields, respectively. The signature of the
metric is $(-+\ldots +)$, $e$ is the gauge coupling and $T$ is the
generic notation for group generators, $D_M=\partial_M+i e T^a A_M^a$.

Suppose that the equations of motion for gravity, scalar and gauge
fields resulting from the action (\ref{action}) admit a solution that
is consistent with four-dimensional  Poincare invariance:
\begin{equation}
ds^2 = G_{MN}dx^M dx^N=\sigma(y)g_{\mu\nu}(x^\mu)dx^\mu dx^\nu +
\gamma_{mn}(y)dy^mdy^n~, \label{conf}
\end{equation}
\begin{equation}
 \Phi = \Phi(y),~~A_\mu=0,~~A_a=A_a(y)~,
\label{field}
\end{equation}
where $x^\mu$ refer to the co-ordinates of the four-dimensional world
and $y^m$ are the extra coordinates.

We would like to know if for small energies the perturbations around
a soliton solution describe four-dimensional excitations
incorporating gravity and all other fields of the standard model.

It is well known that a four-dimensional theory can be constructed
along these lines for Kaluza-Klein (KK) backgrounds (for a review see
\cite{kk}) with $\sigma(y^m)=1,~ \Phi(y^m) = A_a(y^m) = 0$ and a
smooth compact manifold describing extra dimensions. A very
attractive feature of this idea is that the gauge and scalar fields
can have completely geometrical origin.  The problem with this
approach is that fermions are vector-like \cite{Witten:1983ux} and
not chiral, as required by the standard model.

For KK reductions, but now with gauge and scalar backgrounds, the
problem of chiral fermions can be overcome and semi-realistic models
can be constructed. One of the examples is a six-dimensional model
with geometry $M_4\times S_2$, containing in addition to gravity a
U(1) gauge field that has a monopole configuration on $S_2$. The
presence of the U(1) field is essential for the chiral character of
4-dimensional fermions
\cite{Randjbar-Daemi:1982hi,Randjbar-Daemi:1983bw}. More complicated
models incorporating instanton backgrounds can also be constructed
\cite{Randjbar-Daemi:1983qa,Randjbar-Daemi:1983qb}. The Calabi-Yau
compactification in string theory or their orbifold limit
\cite{GSW:1987} are more sophisticated examples of this type.

In these examples the zero modes of all the fields are separated
from other excitations by a mass gap, and the resulting low energy
effective theory is indeed 4-dimensional.  The reason for the
existence of the mass gap is the compact character of extra
dimensions.

It would be interesting to understand whether compact character of
extra dimensions is a necessary requirement for construction of a
(semi) realistic theory, incorporating gravity and other fields.
There are some partial answers to this question. In a theory without
gravity a background of a topological defect (domain wall formed by a
scalar field in 5d, Nielsen-Olesen string in 6d, monopole in 7d,
etc.) may confine chiral fermions and scalar fields and thus give an
effective theory in 4d containing fermions and scalars
\cite{Rubakov:bb,Akama:jy}. As in KK case, the spectrum of
excitations contains a mass gap that insures the effective 4d
character of low-lying excitations, but  contrary to the KK
compactification, the spectrum of higher lying modes is continuous
rather than discrete.

A combination of a topological defect and KK ideas leads to an
exciting possibility of large extra dimensions 
\cite{Arkani-Hamed:1998rs} (see also \cite{Antoniadis:1990ew}), where
standard model fields are assumed to be  localized to a brane
\cite{Polchinski:1995mt}, whereas gravity lives in a bulk.

If gravity is included, there are backgrounds that lead to an
acceptable effective theory of gravity in 4 dimensions
\cite{Randall:1999vf}. They include a non-trivial warp factor
\cite{Rubakov:1983bz} $\sigma(x^a)$ and the presence of a brane,
which in the field theory language is nothing but some topological
defect residing in a higher-dimensional space-time. In 5 dimensions
\cite{Randall:1999vf} or in spherical symmetric case of higher
dimensions
\cite{Gherghetta:2000qi,Gherghetta:2000jf,Giovannini:2001hh} the
behavior of the warp factor is taken to be exponential, $\sigma
=\exp(-c r)$, where $r$ is a radial coordinate. A qualitative
difference between KK case is that the spectrum of graviton
excitations is continuous, and starts from zero without any mass gap.
Still, the theory is acceptable as the interaction of bulk gravitons
with other fields living on the brane is weak at small energies (see,
for a review \cite{Rubakov:2001kp} and references therein).

A first step in constructing a low energy effective theory is an
analysis of the spectrum of small perturbations around the solution.
Consider the spectrum of small fluctuations of the metric, gauge and
scalar fields and fermionic excitations near background
(\ref{conf},\ref{field}). An effective 4-dimensional low-energy
theory could arise if the following conditions are satisfied:

(i) The spectrum contains normalizable zero (or small mass) modes of
graviton, gauge, scalar and fermion fields, with wave functions of
the type $exp(i p_\mu x^\mu) \psi(y^m)$.

(ii) The effects of higher modes should be experimentally
unobservable at low energies, i.e. there should be a mass gap between
the zero modes and excited states. Another possibility is that extra,
unwanted modes may be light but interact very weakly with the zero
modes.

In fact, these conditions are by no means necessary for the existence
of an acceptable 4-dimensional effective theory. For example, in
\cite{Dubovsky:2000am}, a five-dimensional theory without any
normalizable modes was considered, still leading to an effective
theory in 4 dimensions for metastable states localized on a domain
wall. In \cite{Dvali:1996xe,Dvali:1996bg} it was argued that in spite
of the fact that perturbative treatment of gauge field does not
reveal any localized massless modes on domain wall the confinement
effects may lead to the trapping of massless gauge field on a brane.
In \cite{Dvali:2000rx} it was proposed that the fermionic zero modes
localized on a brane may modify through radiative corrections the
effective action for the gauge fields, leading to effectively
four-dimensional interactions at low energies (see, however,
\cite{Dubovsky:2001pe}).

The aim of the present paper is to analyze the small fluctuations of
different fields for a general case of a field-theoretical brane.
Only bosonic fluctuations will be considered; for some partial
results concerning fermionic excitations see, e.g.
\cite{Randjbar-Daemi:2000cr}.

To our  knowledge, this problem has never been addressed in its full
generality, though there are many results for a number of specific
models. The most studied sector is the one related to spin-two fields
(graviton). Its dynamics happens to be related to the background
geometry only, as was shown already in \cite{Randall:1999vf}. A
number of works were devoted to the problem of vector fluctuations in
Randall-Sundrum (RS) model and in a model of Nielsen-Olesen string in
six dimensions \cite{Oda:2000zc}-\cite{Giovannini:2002jf}. The problem of scalar fluctuations of
field-theoretical branes, important for the analysis of their
stability and of the possibility of spontaneous symmetry breaking has
attracted much less attention because of its complexity (see, e.g.
\cite{Giovannini:2001fh}.)

The approach we choose in our paper allows to study all types of
fluctuations on the same basis. The use of the light-cone gauge
allows to separate easily fluctuations of different spins, the use of
the action instead of equations of motion simplifies a lot of the
calculations. What concerns the problem of spin-two fields, we add
nothing new here and present the results just for completeness. As
for the spin-one case, the main difficulty here is the mixing between
the gauge fields and the metric. So, we present the general analysis
of vector fluctuations for an arbitrary field theory  and apply the
formalism to the studied cases of the RS model  and to the case of 
string, reproducing the known results in a much simpler and
transparent way. As a new application, we consider the case of a
monopole configuration.  The spin-zero fluctuations represent the
main difficulty because of the mixing between the components of
scalar field with the components of the metric and the gauge field.
The action for the scalar components, derived in this paper, has
quite a formidable form presented in eqs. (43-48). In section 5 we
clarify the question of scalar physical degrees of freedom in the RS
modes which was obscure in the previous studies.

The paper is organized as follows. In Section 2 we consider a general
decomposition of perturbations on tensors, vectors and scalars and
define variables that allow independent treatment of excitations of
different spins. In  Section 3 we consider application of this
general formalism to the metric of the type
\begin{equation}
ds^2 =
e^{A(r)}\eta_{\mu\nu}(x^\mu)dx^\mu dx^\nu + e^{B(r)} g_{mn}(y)dy^m
dy^n~+dr^2,
\label{simpl}
\end{equation}
where $\eta_{\mu\nu}$ is a flat metric, and warp factors $A$ and $B$
depend on one coordinate ($r$) only.  In Section 4 we show how to
separate fluctuations of higher spins from lower spins in this gauge
with the use of Lorentz invariance on the brane. In section 5, we
apply the formalism to the Randall-Sundrum model. In section 6 we
consider a number of specific models of compactifications on
topological defects (local string in 6d, abelian and non-abelian
monopole in 7d, etc.)  and address the question whether an acceptable
gauge theory can arise from extra dimensions. We show that the string
solution in any number of dimensions leads to normalizable localized
massless vector fields while the monopole solution does so only for
$D_1\geq 4$. Section 7 contains concluding remarks. In Appendix A we
give the components of the curvature tensors for the warped metric.
In Appendix B we give the bilinear part of the spin-one action for
the infinite tower of the Kaluza-Klein modes in a monopole
background. In this appendix we show that the presence of the scalar
field which generates the brane (or the branes, if we have more than
one brane) does not directly affect the vector bilinears. This means
that the bilinear part which is valid everywhere in the space time is
formally identical to the one we would have in the bulk except that
the metric functions $A$ and $B$ will be modified in the vicinity of
the branes.

\section{General formalism}
To analyze the spectrum of small bosonic oscillations around a given
background solution of the classical field equations we insert the
decomposition
\begin{equation}
G_{MN}\rightarrow  G_{MN} + h_{MN},\quad\quad
A_M  \rightarrow  A_M + V_M,\quad\quad
\Phi \rightarrow \Phi +\phi
\label{pert}
\end{equation}
in the action and expand in powers of $h_{MN}$, $V_M$ and $\phi$. On
the right-hand side of (\ref{pert}) $G_{MN},~A_M$ and $\Phi$ refer to
the background solution (\ref{conf},\ref{field}).  The zeroth order
term will give us the value of the classical action evaluated at the
background configuration. The first order terms will be absent
because the background configuration satisfies the classical
equations of motion, which are
\begin{equation}
D_M D^M\Phi = \frac{\partial U}{\partial\Phi^\dagger}~,
\label{sca}
\end{equation}
\begin{equation}
D_M F^{MN} = ie\left((D^N \Phi)^{\dagger} T\Phi -
\Phi^{\dagger}TD^N\Phi\right)~,
\label{YM}
\end{equation}
\begin{eqnarray}
R_{MN}- \frac{1}{2}G_{MN}R &=&
 \frac{\kappa^2}{2}\left(F_{MS}F_N^{\quad S} -
\frac{1}{4}G_{MN}F^2
\right.+\\
(D_M \Phi)^{\dagger} D_N\Phi &+&(D_N \Phi)^{\dagger} D_M\Phi -
\nonumber
\left. G_{MN} (D_S\Phi)^{\dagger}  D^S\Phi - G_{MN} U(\Phi)
\frac{}{}\right)~.
\label{Ei}
\end{eqnarray}
We assume that the scalars belong to some representation of the
gauge group with generators $T$.

These equations of motion are quite general and incorporate
Kaluza-Klein compactification for Einstein-Yang-Mills systems
\cite{Randjbar-Daemi:1982hi}-\cite{ Randjbar-Daemi:1983qb}, pure
gravity warp-factor geometries
\cite{Rubakov:1983bz,Randjbar-Daemi:1985wg}, thick domain walls
\cite{DeWolfe:1999cp,Csaki:2000fc}, global
\cite{Cohen:1999ia,Gregory:1999gv,Olasagasti:2000gx} and local
\cite{Gherghetta:2000qi,Giovannini:2001hh} topological defects in
higher dimensions, etc.

The bilinear in perturbation parts can be written as a sum of several
terms which consist of pure gravitational, gauge and scalar
fluctuations plus terms which contain their mixings, viz,
\begin{equation}
S_2 = S_2(h,h) + S_2(V,V) + S_2(\phi,\phi) + S_2(h,V) +
S_2(h,\phi) + S_2(V,\phi)~.
\label{S_2}
\end{equation}

Each individual term is given by
\begin{eqnarray}
S_2(h,h)=\int d^D X \sqrt{-G}\left\{\frac{1}{2\kappa^2}
\left[\left(h^{ML}_{;M}
-\frac{1}{2}h^{;L}\right)^2
-\frac{1}{2}h^{KS}_{;M}h_{KS}^{;M} + \frac{1}{4}h^{;M}h_{;M}\right]
\right.\\
- \frac{1}{4\kappa^2}R_1 h^2
\nonumber
\label{h}
 - \frac{1}{2}h_{KM}h^K _N \left(\frac{1}{2}F^{MS}F^N_{\quad S} +
(D^M\Phi)^{\dagger}D^N\Phi\right)\\
\nonumber
\left.\hspace{1.5cm}-\frac{1}{2}h^{MN}h^{KS}
\left(\frac{1}{\kappa^2}R _{KMNS}-
\frac{1}{2}F_{KM}F_{NS}\right)\right\}~,
\end{eqnarray}
\begin{eqnarray}
S_2(V,V)= \int d^D X \sqrt{-G}\left\{-\frac{1}{2}D_M V_N D^M V^N
+\frac{1}{2} (D_M V^M)^2 - \frac{1}{2}R^{MN}V_M V_N\right.\\
\left.- e F_{MN}V^M\times V^N \nonumber  -
\frac{e^2}{2}G^{MN} V^a _M
V^b _N \Phi^{\dagger}\{T^a, T^b\}\Phi \right\} ~,
\label{v}
\end{eqnarray}
\begin{equation}
S_2(\phi,\phi)=-\int d^D X \sqrt{-G}
\left\{(D_M\phi)^{\dagger} D^M \phi
+\frac{1}{2}\phi \frac{\partial^2 U}{\partial \Phi^2}\phi
+\frac{1}{2}\phi^* \frac{\partial^2 U}{\partial {\Phi^*}^2}\phi^*
+\phi^*
\frac{\partial^2 U}{\partial \Phi\partial \Phi^*}\phi\right\}~,
\label{scal2}
\end{equation}
\begin{eqnarray}
S_2(h,V)= -\int d^D X \sqrt{-G}\left\{\left(D^M V^N - D^N V^M\right)
\left(\frac{1}{4}h F_{MN} + h_{LN} F^L_{\quad M}\right)\right.\\
\nonumber
\left. +ie((D_N\Phi)^{ \dagger}T^a\Phi -\Phi^{\dagger}T^a
D_N\Phi)\left(\frac{1}{2}G^{MN} h -h^{MN}\right)V^a_M\right\}~,
 \label{h,v}
\end{eqnarray}
\begin{eqnarray}
S_2(V,\phi)&=& -ie\int d^D X \sqrt{-G}V^M _a
\left\{(D_M\phi)^{\dagger} T^a
\Phi +(D_M\Phi)^{\dagger} T^a\phi\right.\\
\nonumber
& & \hspace{2cm}\left. -\phi^{\dagger}T^a D_M\Phi
-\Phi^{\dagger}T^a D_M\phi\right\}~,
 \label{phi,v}
\end{eqnarray}
\begin{eqnarray}
S_2(\phi,h)&=& \int d^D X \sqrt{-G}\left\{
\left(h^{MN}-\frac{1}{2}G^{MN}h\right)
\left((D_M\phi)^{\dagger}D_N\Phi +
(D_M\Phi)^{\dagger}D_N\phi\right)\right.\\
\nonumber
& &\left.
 \hspace{2cm}+\frac{1}{2}\phi h\frac{\partial U}{\partial \Phi}
+\frac{1}{2}\phi^* h\frac{\partial U}{\partial \Phi^*}
\right\}~,
\label{phi,h}
\end{eqnarray}
where $h= G^{MN}h_{MN}$, and
\begin{equation}
\frac{2}{\kappa^2}R_1= \frac{1}{\kappa^2}R  -\frac{1}{4}F^2 -
D_M\Phi^{\dagger}D^M\Phi -U(\Phi)
\label{L}
\end{equation}
denotes the value of the classical Lagrangian evaluated at the
background solution. In these expressions the covariant derivatives
contain the background gravitational as well as the gauge
connections. We have made use of the classical equations of motion to
simplify the expressions.

The bilinear action should inherit the linearized gauge symmetries of
the original Einstein, Yang-Mills, scalar system. The linearized
general coordinate transformations are given by
\begin{equation}
\delta_{\xi} h_{MN}= -\xi_{N;M}-\xi_{M;N}~,
\label{gc1}\
\end{equation}
\begin{equation}
\delta_{\xi} V_{M}= -\xi^{L}F_{LM}-D_M \chi~,
\label{gc2}\
\end{equation}
\begin{equation}
\delta_{\xi} \phi= -\xi^{M}D_M\Phi + i e  \chi\Phi~, 
\label{gc3}
\end{equation}
where $\xi$ is the infinitesimal parameter of the general coordinate
transformations and $\chi$ is the infinitesimal  parameter of an
induced Yang-Mills gauge transformation defined by
\[
\chi= \xi^L A_L~.
\]

We can write down similar transformation rules under the Yang-Mills
gauge transformations. Because of such local symmetries, if we couple
$h_{MN}$, $V_{M}$ and $\phi$ to external sources via
\begin{equation}
S_{int} =\int d^D X \sqrt{-G}[\frac{1}{2}h^{MN}T_{MN} +
J^MV_M+J \phi]
\label{coup}
\end{equation}
the source terms must satisfy the following conservation
laws,
\begin{equation}
D_M T^{MN}= J^M F^N_{\quad M} + J D^N\Phi~,
\label{con1}
\end{equation}
\begin{equation}
D_M J^M= ie J\Phi~.
 \label{con2}
\end{equation}
These gauge symmetries can be used to simplify the analysis of the
spectrum of small oscillations. The most frequently used gauges in
the literature on Kaluza Klein theory and on supergravity theories
have been the covariant gauges. For the analysis of the next section
we shall find it more convenient to use the light-cone gauge.

\section{Warped geometries}
The bilinear action of the previous section is completely general and
can be used to analyze the spectrum of the small oscillations of the
Einstein-Yang-Mills-scalar system around any solution of the
background equations. In this section we shall specialize to a set of
particular solutions, the so called warped solutions which have
attracted considerable amount of interest in recent years. The space
time geometry of a warped solution is characterized by the metric
(\ref{simpl}), where the coordinates $x^\mu$, $\mu=0, 1,....D_1-1$
parameterize a flat Minkowski space with the usual metric tensor
$\eta_{\mu\nu}= diag (-1, +1, ....,+1)$, while the coordinates $y^m$,
$m= 1,..., D_2$ cover a compact $D_2$ -dimensional manifold. A
coordinate $r$ can be thought of as a radial variable. The total
number of the space time dimensions is thus $D=D_1 + D_2 + 1$.

The presence of the standard Poincare symmetry in the subspace
spanned by the  $x$'s enable us to use the representations of this
group to define the usual notions of particle physics. In addition,
if the space covered by the $y$'s also do posses  isometries which
leave the entire classical background invariant, there will be
massless spin 1 particles in the $x$-subspace corresponding to the
usual Kaluza Klein gauge fields. In general the final gauge fields
are linear combinations of the appropriate modes contained in the
fluctuations $h_{m\mu}$ and $V_\mu$. We shall see examples of this
later on.

Another advantage of the Poincare symmetry is that it allows us to
use the light cone gauge to analyze the spectrum.  As we shall see in
this gauge the fields of different spins are readily separated from
each other and, after we have eliminated the dependent and non
physical degrees of freedom, we can read the spectrum almost
directly.

To use the light cone gauge first we introduce the light cone
coordinates $x^{\pm}=\frac{1}{\sqrt{2}}(x^{D_1-1} \pm x^0)$ in the
$x$-subspace. The inner product of any two vectors then become
\begin{equation}
G^{MN} A_M B_N= e^{-A(r)}(A_{+}B_{-} + A_{-}B_{+} + A_i B_i) +
e^{-B(r)}g^{mn}A_m B_n + A_r B_r~,
\end{equation}
where $A_i$ and $B_i$ are the transverse components of $A_M$ and
$B_M$,  respectively, $A_{\pm}=\frac{1}{\sqrt{2}}(A_{D_1-1} \pm
A_0)$. The light cone gauge is defined by
\begin{equation}
V_{-}=0 \quad\quad\quad and \quad\quad\quad h_{-M} = 0,
 \quad\quad for\quad all \quad M~.
\end{equation}
Using these gauge conditions it turns out that the $+$ components are
not independent but can be expressed in terms of the transverse
objects $V_i$ and $h_{ij}$. Since the calculations leading to these
results are  simple but rather lengthy here we give the results only
\footnote{This is very similar to the light cone formulation in
superstring theory, except that in our field theory model the
calculations leading to these results are somewhat more lengthy. For
the application of the light cone gauge in the context of the higher
dimensional theories of gravity see 
\cite{Randjbar-Daemi:1984ap,Randjbar-Daemi:1984fs}.}.

First in the Yang - Mills sector we obtain the constraint equation
\begin{equation}
\partial_- V_+ = - \partial_i V_i - e^A \left(\frac{D_1 -2}{2} A' V_r
+ D_{\underline m} V^{\underline m}\right) -ie^A
\left(\phi^{\dagger}T\Phi - \Phi^{\dagger}T\phi\right)e~,
\label{v+}
\end{equation}
where the underlined index $\underline m$ stands for the pair of
indices $m$ , which is tangent to the $y$-space and $r$. In the
Einstein sector the $h_{++}$ field equation simply leads to
\begin{equation}
h=0~.
\end{equation}
This brings about considerable amount  of simplification. Next after
some considerable amount of calculations  we obtain the constraint
equations for $h_{i+}$, $h_{m+}$ and $h_{r+}$. First consider the one
for $h_{i+}$,
\begin{equation}
\partial_{-} h_{i+}= - \partial_{j} h_{ij}  -e^A\left(\frac{D_1-1}{2}
A' h_{ir}  + h_{i\underline l ;}^{\quad\underline l}\right)~.
\label{h+1}
\end{equation}
Next consider the constraint equation for $h_{m+}$,
\begin{equation}
\partial_{-} h_{m+}= - \partial_{i} h_{im} -e^A \left(\frac{D_1
-2}{2} A' h_{rm} + h_{m\underline l ;}^{\quad\underline l}
-\kappa^2\left(\phi^{\dagger} D_m\Phi + (D_m
 \Phi)^{\dagger}\phi + V_{\underline l}F_n^{\quad\underline
 l}\right)\right)~.
 \label{h+2}
\end{equation}
Finally we have the constraint equation for  $h_{r+}$, which reads
as follows,
\begin{equation}
\partial_{-} h_{r+}= - \partial_{i} h_{ir}  + \frac{A'}{2} h_{ii}-e^A
\left(\frac{D_1 -2}{2} A' h_{rr} + h_{r\underline l
;}^{\quad\underline l} -\kappa^2\left(\phi^{\dagger} D_r\Phi + (D_r
\Phi)^{\dagger}\phi+ V_{\underline l}F_r^{\quad\underline
l}\right)\right)~.
 \label{h+3}
\end{equation}
There is also a constraint equation for $h_{++}$ component of the
metric, which is quite complicated. We will not write it as it is not
used in the analysis to follow.

It is possible to write the above constraint equations in a more
compact form. For the gauge field constraint the compact form reads
as
\begin{equation}
D_MV^M= -i \left(\phi^{\dagger}T\Phi -
\Phi^{\dagger}T\phi\right)e~,
\end{equation}
 while for the gravitational perturbation the compact form becomes,
\begin{equation}
D_M h^{M}_{N} = \kappa^2\left(\phi^{\dagger} D_N\Phi +
(D_N \Phi)^{\dagger}\phi+ V_{\underline l}F_N^{\quad\underline
l}\right)~.
\end{equation}
Inserting our background configuration in these equations and
expanding them in components reproduces the above gauge and
gravitational constraint equations.

Using these constraint equations together with the light cone gauge
conditions enable us to separate the bilinear Lagrangian into various
spin sectors. The advantage of the light cone gauge is that the
quadratic Lagrangian for the  fields of different spins disentangle
from each other.

\subsection{Spin-2 Action}
Using the constraints and the gauge conditions in the general
bilinear action given in section two the spin-2 part separates from
the rest and its  action takes the following the form:
\begin{eqnarray}
 S(spin-2)= -\frac{1}{4\kappa^2}\int d^D X \sqrt{-G}\left[
\partial_\mu \tilde h_{ij}\partial^{\mu} \tilde h^{ij} + {A'}^2
\tilde h^{ij}\tilde h_{ij} + G^{ik}G^{jl}\partial_r \tilde
h_{ij}\partial_r \tilde h_{kl} \right.\\
\nonumber
\left. - 2 A' \tilde h^{ij}\partial_r \tilde h_{ij} + D_m \tilde
h_{ij} D^m \tilde h^{ij} \right]~.
\label{spin2}
\end{eqnarray}
Here and in the following all the  indices are raised and lowered
with the $r$-dependent metric $G^{\mu\nu}= e^{-A(r)} \eta ^{\mu\nu}$
and $G^{mn}= e^{-B(r)}g^{mn}(y)$,  $\tilde h_{ij}$ indicates the
traceless part of $h_{ij}$, viz,
\begin{equation}
h_{ij} = \tilde h_{ij} + \frac{1}{D_1-2}G_{ij}h_k ^{k}~,
\end{equation}
where $ h_k ^{k} = G^{ik}h_{ik}$.

It is to be observed that the spin-two quadratic action is quite
universal. Its dependence on the background scalar and Yang- Mills
fields is only through the dependence of the space time geometry on
these fields. In other words, the background
scalars and gauge fields do not directly enter into the spin-2
action. This is not true for the spin-1 and spin-0 quadratic actions.

This action has more transparent form in terms of the field with one
upper and one lower index:
\begin{equation}
 S(spin-2)= -\frac{1}{4\kappa^2}\int d^D X \sqrt{-G} \left[e^{-A}
\eta^{\mu\nu}\partial_\mu\tilde h_i^j\partial_{\nu}\tilde h^i_j  +
\partial_r \tilde h_i^j \partial_r \tilde h_j^i  + D_m \tilde h_i^j
D^m \tilde h^i_j\right]~.
\label{s2better}
\end{equation}

\subsection{Localized spin-2 Fields}
Since $\tilde h_i ^{j}$ is a scalar field in the $y$ subspace,
provided this space is compact, the operator $D_m$ will have a
normalizable zero mode.  We can take zero mode to be $r$ independent,
viz\footnote{There is also a $r$-dependent zero mode which will be
discussed in section 6 in analyzing the localization of the spin-1
fields.},
\begin{equation}
\partial_r \tilde h_i ^{j} =0~.
\end{equation}
A little calculation will then show that the $S(spin-2)$ reduces to
\begin{equation}
S(spin-2)= -\frac{1}{4}\int d^D X \sqrt{-G}e^{-A}\eta^{\mu\nu}
\partial_\mu
\tilde h_i^j\partial_{\nu} \tilde h^i_j~.
\label{spin2z}
\end{equation}
Upon integration over the $D_2$ dimensional $y$ space, which is
trivial as the fields are independent of $y$ and integrating over the
$r$ subspace, which in contrast is non-trivial, we obviously obtain
the light cone gauge action for a graviton in the $D_1$ dimensional
subspace covered by the $x$ coordinates. This action will be
meaningful provided the $r$ integral is finite. The condition for the
finiteness of the $r$-integral is
\begin{equation}
\int dr e^{\frac{D_1-2}{2} A(r)+ \frac{D_2}{2}B(r) }<\infty~,
\end{equation}
which is nothing but the condition of the finiteness of the
$D_1$-dimensional Planck scale \cite{Randall:1999vf}.

\subsection{Spin-1 Action}
The basic spin-1 fields are $V_i$ and $h_{\underline m i}$. In the
bilinear part of the action, in general, these fields mix with each
other as well as with the derivatives of the scalar fields. However,
in the light cone gauge there is a remarkable simplification due to
the fact that there are no mixing with the fluctuations of spin zero
fields. The bilinear action is given by,
\begin{equation}
S(spin-1)= \int d^D X \sqrt{-G}\left(L_v(V,V)+ L_v(h,h)
+ L_v(h,V)\right)~,
\label{spin1}
\end{equation}
where the different pieces are:
\begin{eqnarray}
\nonumber L_v(V,V) =  - \frac{1}{2} \left(\partial_\mu
V_i\partial^{\mu} V^i + e^{-A} \partial_r V_i \partial_r V_i + D_m
V_i D^m V^i + e^2 V_i ^a
V^{ib}\Phi^{\dagger} \{T^a,T^b\}\Phi\right)~,\\
\end{eqnarray}
\begin{eqnarray}
L_v(h,V) = - D^{\underline m}V_i h^{\underline l i}F_{\underline
l\underline m} -\frac{1}{2}A'  V_i h^{\underline l i}
F_{\underline l r} +ie ((D_{\underline n}\Phi)^{\dagger}T\Phi -
\Phi^{\dagger}TD_{\underline n}\Phi) h^{\underline n i}V_i~,
\end{eqnarray}
\begin{eqnarray}
\nonumber
L_v(h,h) = -\frac{1}{2\kappa^2}\left(\partial_\mu h_{i\underline
m}\partial^\mu h^{i\underline m} + \partial_r h_{i\underline
m}\partial^r h^{i\underline m} + D_n h_{i\underline m} D^n
h^{i\underline m}\right)\\
\nonumber
 - \frac{1}{4\kappa^2} h_{mi} h^{mi}\left(\frac{1}{2}{B'}^2 + {A'}^2
 \right)
 - \frac{1}{4\kappa^2} h_{ri} h^{ri}\left(D_2 A'B'
 -\frac{D_1-2}{2}{A'}^2 \right)
+  \frac{1}{\kappa^2} {A'} h^{ir} D^m h_{i m}\\
 - \frac{1}{2} h_{\underline m i} h_{\underline n} ^i
\left(\frac{1}{2} F^{\underline m}_{\quad \underline k}
F^{\underline n \underline k}
+ (D^{\underline m}\Phi)^{\dagger}D^{\underline n}\Phi\right)~.
 \label{vhh}
\end{eqnarray}
Note that if $F_{r n} = 0$ and $\Phi$ is $r$ independent or does not
couple to the gauge field, the only mixing is between $V_i$ and
$h_{mi}$ and between the divergence $D^m h_{mi}$ and $h_{ri}$. The
seemingly $3\times 3$ system consisting of $V_i$, $h_{mi}$ and
$h_{ri}$ will reduce to two $2\times 2$ system. We shall see an
example of such simplification in the next section.

\subsection{Spin-0 Action}
The spin zero action contains the fields
$h_{\underline m \underline n}$, $V_{\underline m}$, $h^i_i$ and
$\phi$.
It is the most complicated one and has the form:
\begin{equation}
S(spin-0)= \int d^D X \sqrt{-G}\left[L(h,h) + L(V,V) + L(\phi,\phi)
+ L(h,V) + L(h,\phi) +L(V,\phi)\right]~,
\label{spin0}
\end{equation}
where the different pieces are:
\begin{eqnarray}
L(h,h)=-\frac{1}{4\kappa^2}\left\{\frac{}{}
\partial_\mu h_{\underline m \underline n}
\partial^\mu h^{\underline m \underline n}
+\partial_r h_{\underline m \underline n}
\partial^r h^{\underline m \underline n}
+D_l h_{\underline m \underline n}
D^l h^{\underline m \underline n}\right.
\nonumber
\\
+h_{mr}h^{mr}\left(\frac{4-D_1}{2}{A'}^2 +D_2 A' B' + 2 A''
-\frac{3}{2}{B'}^2-2B''\right)
\nonumber
\\
\nonumber
+\frac{1}{2}h_{mn}h^{mn}{B'}^2
+h_{rr}^2\left(\frac{4-D_1}{2}{A'}^2+D_2 A' B'+ 2 A''\right)
\\
-4A'h^{r\underline m}D_n h^n_{\underline m} +
2(A''+{A'}^2){h_r^r}h_i^i + \frac{1}{2} {A'}^2 {h_i^i}^2 +
A'B'h_i^i h_m^m\\
+2(B''+B')h_m^m h^{rr}
+ \frac{1}{2} {B'}^2{h_m^m}^2 - 2
e^{-B}\Omega_{mn}^{\quad kl} h^m_l h^n_k \nonumber
\\
+ 2\kappa^2 h_{\underline m}^{\underline l}h_{\underline l \underline
m} \left(\frac{1}{2}F^{\underline m}_{\quad \underline k}
F^{\underline n \underline k}+(D^{\underline m}\Phi)^\dagger
D^{\underline n}\Phi\right)
-\kappa^2h_{\underline k \underline s}h_{\underline m \underline n}
F^{\underline k \underline m} F^{\underline s \underline n}
\nonumber
\\
\left.
+\frac{1}{D_1-2}\left(\partial_\mu h_i^i\partial^\mu h_j^j
+(\partial_r h_i^i)^2 + D_m h_i^iD^m h_j^j\right)\right\}~,
\nonumber
\end{eqnarray}

\begin{eqnarray}
L(V,V)=-\frac{1}{2}\left(
\partial_\mu V_{\underline m}\partial^\mu V^{\underline m}
+ D_{\underline m}V_{\underline n}D^{\underline m}V^{\underline n}
+\frac{4-3D_1}{4}{A'}^2V_r^2 \right.
\nonumber
\\
- 2 A' V_r D_{\underline m}V^{\underline m}
+R_{\underline m \underline n}V^{\underline m}V^{\underline n}
+2eF_{\underline m \underline n}V^{\underline m}\times V^{\underline
n}
\\
\nonumber
\left.
+e^2 V_{\underline m}^aV^{\underline m b}\Phi\{T^a T^b\}\Phi
+\kappa^2(V^{\underline l}F_{\underline l \underline m})^2\frac{}{}
\right)~,
\end{eqnarray}

\begin{eqnarray}
L(\phi,\phi)= -(D_M\phi)^{\dagger} D^M \phi
-\frac{1}{2}\phi \frac{\partial^2 U}{\partial \Phi^2}\phi
-\frac{1}{2}\phi^* \frac{\partial^2 U}{\partial {\Phi^*}^2}\phi^*
-\phi^* \frac{\partial^2 U}{\partial \Phi\partial \Phi^*}\phi
\nonumber
\\
+\frac{e^2}{2}\left(\phi^\dagger T \Phi - \Phi^\dagger T
\phi\right)^2
-\frac{\kappa^2}{2}\left(\phi^\dagger D_{\underline m} \Phi +
(D_{\underline m} \Phi)^\dagger \phi\right)^2~,
\end{eqnarray}

\begin{equation}
L(h,V)=V^{\underline n}
\left(D_{\underline m}h_{\underline l \underline n}F^{\underline l
\underline m} -h_{\underline l}^{\underline m}D_{\underline m}
F^{\underline l}_{\quad \underline n}\right)
-\frac{1}{2}A' F^{r \underline m}V_{\underline m}h_i^i
+\frac{D_1-2}{2}A' F^{\underline l \underline m}
V_{\underline m}h_{r\underline l}~,
\end{equation}

\begin{eqnarray}
L(h,\phi)=h^{\underline l \underline m}_{\quad;{ \underline l}}
\left(\phi^\dagger D_{\underline m}\Phi + ( D_{\underline
m}\Phi)^\dagger \phi\right)
-\frac{1}{2}h_i^i A' \left(\phi^\dagger D_r\Phi +
(D_r\Phi)^\dagger\phi\right)
\nonumber
\\
+\frac{D_1-2}{2}h^{r \underline m} A'
\left(\phi^\dagger D_{\underline m}\Phi + (D_{\underline
m}\Phi)^\dagger \phi\right) \\
+h^{\underline m\underline
n}(D_{\underline m}\phi^{\dagger} D_{\underline n}\Phi +
D_{\underline m} \Phi^{\dagger}D_{\underline n} \phi)~,
\nonumber
\end{eqnarray}

\begin{eqnarray}
L(V,\phi)=2 i V^{\underline m}
\left(\phi^\dagger T D_{\underline m}\Phi - (D_{\underline
m}\Phi)^\dagger T \phi\right)
-i \frac{D_1-2}{2} A' V_r\left(\phi^\dagger T \Phi - \Phi^\dagger T
\phi\right)
\nonumber\\
-\kappa^2F^{\underline l \underline m}V_{\underline m}
\left(\phi^\dagger D_{\underline l}\Phi + (D_{\underline
l}\Phi)^\dagger \phi\right)~.
\end{eqnarray}

We recall that the three fields $h^m_{m}, h^r _{r}$ and $h^i _{i}$
are not independent as their sum vanishes due to the constraint
$h=0$. The spin-zero action can be used for an analysis of stability
of different brane solutions, see, e.g. \cite{Lavrelashvili:aa}.

\section{Lorentz invariance and the modes decomposition}

The light cone gauge is ideally suited for the study of massless
excitations because it leaves only  physical degrees of freedom for
$D_1$-dimensional graviton and for the vector fields. As for the
massive modes, it distributes the members of the same spin multiplet
over tensor, vector and scalar equations.   A spin two massive
particle has $D_1(D_1-1)/2-1$ degrees of freedom, from which
$D_1(D_1-3)/2$ components are in the gravity multiplet
(\ref{spin2}),  $D_1-2$ components in the vector sector
(\ref{spin1}), and one component in the scalar sector (\ref{spin0}).

By the same reasoning, for massive vector excitations $D_1-2$
components are present in the vector part (\ref{spin1}) while one is
hidden in the scalar sector (\ref{spin0}). This means that one
combination of vector fields and one combination of scalar fields
must obey  equations which give the same spectrum of fluctuation  as
the tensor part. If the total number of physical vector fields is
$N_v$, then $N_v$ combinations of scalar fields can be decoupled from
scalar equations and must have the same massive spectrum, as follows
from the vector equations. In this section we will show how to single
out these combinations.

To study a spectrum of $D_1$-dimensional perturbations it is
convenient to work in Fourier space, with
\begin{equation}
h_{MN} \propto V_{M} \propto  \exp{(i p_{\mu}x^{\mu})}~.
\end{equation}
For the massive modes one can always choose a coordinate (rest frame
of the massive particle) system in which $p_{\mu} = (m,\dots,0)$,
where $m$ is a mass of an excitation. We shall be using this frame in
this Section.

We start from the massive spin-two multiplet. From the
$D_1$-dimensional point of view it is described by a tensor field
$H_{\mu\nu}$ which is transverse and traceless,
\begin{equation}
\partial_\mu H^\mu_\nu=0,~~~
H^\mu_\mu=0~.
\label{4dtens}
\end{equation}
This tensor is invariant under $D_1$-dimensional general-coordinate
transformations.

In the rest frame the conditions (\ref{4dtens}) give $H_{0\mu} =0$
and $\sum_{i=1}^{D_1-1} H_{ii}=0$. The components of the tensor
$H_{\mu\nu}$ automatically satisfying the conditions (\ref{4dtens}) can
be expressed through the arbitrary metric as
$\tilde{h}_{ij}$, then $h_{i(D_1-1)}$~, and, finally
$\sum_{i=1}^{D_1-2} h_{ii}-\frac{D_1-2}{D_1-1}\sum_{i=1}^{D_1-1}
h_{ii}$. The first term is contained in the spin two part of
the action (\ref{spin2}), the second term is counted as a vector
in the light-cone gauge, and the last one as a scalar. The expression
in terms of the fields in this gauge are just $h_{i+}$ for the
vectors, which is converted to
\begin{equation}
\frac{D_1-1}{2} A'h_{ir}  + h_{i\underline l ;}^{\quad\underline l}~.
\label{grvect}
\end{equation}
with the use of constraint (\ref{h+1}). It is this combination of
the vector fields that may be decoupled from the vector equations
(\ref{spin1}) and should have the same spectrum as the tensor
part. The scalar field that can be decoupled from (\ref{spin0})
and has the same sect rum as (\ref{spin2}) is given by
\begin{equation}
h^i_i+(D_1-2)h^+_+~,
\label{grsc}
\end{equation}
where $h_{++}$ has to be expressed through physical scalar components
via constraint equation, which we do not write for a general case
because of its complexity (see, however, below).

A similar line of reasoning can be applied for the massive vector
fields. From the four-dimensional point of view they are described
by the vector fields  $h_{\underline m\nu}$ and $V_\mu$ which are
transverse,
\begin{equation}
\partial_\mu h^\mu_{\underline n}=0,~~ \partial_\mu V^\mu~=0.
\label{4vect}
\end{equation}
In the rest frame we have $h^0_{\underline n}=0$ and $V^0=0$.
Thus, the components $h_{+\underline n}$ and $V_+$, expressed
through the constraints (\ref{v+}), (\ref{h+2}) and (\ref{h+3})
should give the desired combinations of scalar fields, that have
the same spectrum as the corresponding vector modes:
\begin{equation}
\frac{D_1 -2}{2} A' V_r +
D_{\underline m} V^{\underline m} -i \left(\phi^{\dagger}T\Phi -
\Phi^{\dagger}T\phi\right)e~,
\label{vs0}
\end{equation}
\begin{equation}
\frac{D_1
-2}{2} A' h_{rm} + h_{m\underline l ;}^{\quad\underline l}
-\kappa^2\left(\phi^{\dagger} D_m\Phi + (D_m
 \Phi)^{\dagger}\phi + V_{\underline l}F_n^{\quad\underline
 l}\right)~,
 \label{vs1}
\end{equation}
\begin{equation}
 \frac{A'}{2} h^i_i-
\left(\frac{D_1 -2}{2} A' h_{rr} + h_{r\underline l
;}^{\quad\underline l} -\kappa^2\left(\phi^{\dagger} D_r\Phi +
(D_r \Phi)^{\dagger}\phi+ V_{\underline l}F_r^{\quad\underline
l}\right)\right)~.
 \label{vs2}
\end{equation}

\section{The  Randall-Sundrum model}

The simplest model where the general equations presented above can be
applied is the Randall-Sundrum model in which a thick domain  wall is
formed by a real scalar field $\Phi$ with potential $U(\Phi)$. This
model has been studied in many papers. Here  we just present some
peculiarities that are present in the high-cone gauge. For this model
$D_1=4,~D_2=0$. However, we shall write the equations in some
generality which will also be useful for the study of the monopole
background in the next section. For arbitrary $D_1$ and $D_2$ we choose
the following anzatz for the gauge and the real scalar field,
\begin{equation}
 \Phi = \Phi(r),~~A_\mu= A_r=0,~~A_m=A_m(y)~,
\label{fields}
\end{equation}
where $x_\mu$, $\mu= 0, 1, ..., D_1-1$ refer to the co-ordinates of
the $D_1$-dimensional world and $y^m$, $m=1,..., D_2$ and $ r $ are
the extra coordinates.

Inserting our anzatz into the bosonic field equations we obtain,
\begin{eqnarray}
\frac{D_1}{4}(D_1-1)A^{'2} + (D_1-1)A'' +\frac{D_2}{2}(D_1-1)A'B'
\nonumber
\\
+\frac{1}{4}D_2(D_2+1)B^{'2} + D_2B'' -e^{-B}\Omega=\kappa^2 L~,
\label{Ei1}
\end{eqnarray}
\begin{equation}
\frac{D_1}{4}(D_1-1)A^{'2} + \frac{1}{2}D_1D_2A'B'
+\frac{D_2}{4}(D_2-1)B^{'2} -e^{-B}\Omega=\kappa^2(\Phi^{'2}+L)~,
\label{Ei2}
\end{equation}
\begin{eqnarray}
\frac{D_2}{4}(D_2-1)B^{'2} + (D_ 2-1)B'' +\frac{D_1}{2}(D_2-1)A'B' +
\frac{1}{4}D_1(D_1+1)A^{'2} + D_1A''
\nonumber
\\
+(\frac{1}{D_2}-
\frac{1}{2}) e^{-B}\Omega=
 \kappa^2\left(\frac{ e^{-2B}}{e^2}f^2 +L\right)~,
\label{Ei3}
\end{eqnarray}
where $f^2$ in  eq.(\ref{Ei3}) is defined by $g^{mn}F_{mk}F_{nl}= f^2
g_{mk}$, $L=-\frac{1}{4}e^{-2B}f^2 - \frac{1}{2}\Phi^{'2}-U(\Phi)$.
With this assumption we are anticipating that our background solution
is going to be of the same type as the ones already analyzed in
\cite{Randjbar-Daemi:2000ft}.

The Klein-Gordon equation for $\Phi$ can be written as
\begin{equation}
\partial_r (\frac{1}{2}\Phi^{'2} -U) = -(\frac{D_1}{2}A' +
\frac{D_2}{2}B')\Phi^{'2}~.
\end{equation}
We have four equations for three functions $A, B$ and $\Phi$. One can
however, show that the equations are not over determined and they are
consistent. These general equations can be employed to  analyze the
spectrum of many models in detail. We shall use them in the next
section for the case of a monopole background.

Here we specialize to the simple case of the Randall Sundrum model
for which $D_1=4$ and $D_2=0$. The gauge field is also set to zero.
The equation of motion for the scalar field then becomes:
\begin{equation}
\Phi'' + 2 A'\Phi' = \frac{\partial U}{\partial\Phi}~,
\label{scrs}
\end{equation}
whereas Einstein equations can be written in the form
\begin{eqnarray}
A''+{A'}^2=-\frac{1}{3}\kappa^2\left(
\frac{1}{2}{\Phi'}^2 +U \right)~,\\
{A'}^2=\frac{1}{3}\kappa^2\left(
\frac{1}{2}{\Phi'}^2 -U \right)~.
\label{einrs}
\end{eqnarray}
The physical fields are $h_{12}$ and $h_{11}-h_{22}$ (gravity
multiplet), $h_{ri}$ (vector multiplet or graviphoton) and two scalar
fields  $h^i_i$ and $\phi$.

The spin-two Lagrangian has a simple form
\begin{equation}
 L(spin-2)= -\frac{1}{4\kappa^2}\int d^D X \sqrt{-G}
\left(e^{-A}\eta^{\mu\nu}\partial_\mu\tilde h_i^j\partial_{\nu}\tilde
h^i_j  + \partial_r \tilde h_i^j \partial_r
\tilde h_j^i\right)~.
 \label{spin2rs}
\end{equation}
The equation for determination of four-dimensional spectrum is
\begin{equation}
-e^{-2A}\partial_r\left(e^{2A}\partial_r\Psi\right)=m^2e^{-A}\Psi
\label{spectr2rs}
\end{equation}
and allows a normalizable zero mode (graviton) provided $A \to \infty$
for $r \to \pm \infty$ which is achieved if the fine-tuning condition
of \cite{Randall:1999vf} is satisfied.

The spin-one action is
 \begin{equation}
L(spin-1) =
 -\frac{1}{2\kappa^2}\left(\partial_\mu h_{ir}\partial^\mu
 h^{ir} + \partial_r h_{ir}\partial^r h^{ir} \right)
 + \frac{1}{4\kappa^2} h_{ri} h^{ri}{A'}^2
  - \frac{1}{4} h_{r i} h^{ri} \Phi'{^2}~
 \label{spin1rs}
\end{equation}
and can be expressed entirely through the metric with the use of
equation $\frac{\kappa^2}{3}(\Phi')^2 +A''=0$ following from (\ref{einrs}).
The corresponding equation for the determination of the spectrum
\begin{equation}
-e^{-A}\partial_r\left(e^{A}\partial_r\Psi\right)-
\frac{1}{2}\left(2A'' +{A'}^2\right)=m^2e^{-A}\Psi
\label{spectr1}
\end{equation}
is a partner equation of (\ref{spectr2rs}) from the point of view of
supersymmetric quantum mechanics \cite{Cooper:1994eh}. It has exactly
the same spectrum as for spin-two excitations with removed zero mode
-- a result which one would expect from four-dimensional Poincare
invariance on the brane (five components of a four-dimensional
massive spin two field are distributed among the multiplets in the
light-cone gauge as $\tilde h_i^j$ (two degrees of freedom), $h_{ri}$
(two degrees of freedom), with fifth degree of freedom hiding in some
combination of the two scalar fields $\phi$ and $h_{rr}=-h^i_i$.

Finally, the spin-zero action is
\begin{eqnarray}
\nonumber
S(spin-0) = \int d^4x dr e^{2A}\left[-\frac{3}{8}
\left(e^{-A}\eta^{\mu\nu}\partial_\mu h_{rr}\partial_\nu h_{rr}+
(h_{rr}')^2-h_{rr}^2(2A''+{A'}^2)\right)\right.\\
-\frac{1}{2}\left(e^{-A}\eta^{\mu\nu}\partial_\mu\phi\partial_\nu\phi+
(\phi')^2+\frac{\partial^2 U}{\partial^2 \Phi^2}\phi^2+
\phi^2(\Phi')^2\right) \\
\left. -\phi h_{rr}\left(\Phi'' + \frac{1}{2}A'\Phi\right)\right]~.
\nonumber
\label{spin0rs}
\end{eqnarray}
It contains fluctuations belonging to the four dimensional  massive
spin-2 multiplet and a genuine scalar field. A component
corresponding to the spin-2 field can be found in a way  discussed in
the previous section, see eq.(\ref{grsc}). In the particle rest-frame
it is
\begin{equation}
-h_{rr}+2 h^+_+~,
\end{equation}
where
\begin{equation}
h^+_+=-h_{rr}+\frac{e^A}{2m^2}\left[3 h_{rr}({A'}^2-A'') + 3 A'h'{rr}+
2 \phi(\Phi'' - A' \Phi') - 2 \phi' \Phi' \right]~,
\end{equation}
and $m$ is the particle mass.

Another independent component, corresponding to a genuine scalar
field in four dimensions can be found from the requirement of gauge
invariance with respect to general-coordinate transformations as in
ref. \cite{Giovannini:2001fh} and is equal to
\begin{equation}
\phi +\frac{\Phi'}{3A'}(h_{rr}+h^+_+)~.
\label{scalar}
\end{equation}
According to \cite{Giovannini:2001fh}, there are no normalizable zero
modes in the scalar sector.

\section{Vector fluctuations in specific models}
The question of gauge field localization by gravity has been
addressed in a number of papers, see, e.g.
\cite{Oda:2000zc}-\cite{Giovannini:2002jf}. It has been shown that
gauge fields cannot be localized in five-dimensional space time, but
a normalizable zero mode can exist for $D_1=4$, $D_2 \geq 1$. The
spectrum of vector field excitation does not have a mass gap, but
this may not be dangerous from phenomenological point of view because
of the screening phenomenon discussed in \cite{Dubovsky:2002xv}.

In this section we shall consider some specific models in which a
defect is formed by interacting gauge and scalar field. Here
analysis of gauge field localization happens to be quite
complicated because of the mixing of the gauge fields coming from
the metric with the gauge fields. Nevertheless due to the complete
separation of fields of different spins in the light cone gauge
the analyzes of the physical spectrum in our formalism  becomes
considerably simpler than in other gauges.

\subsection{Nielsen-Olesen string in six dimensions}
As we have seen above the minimal Randall-Sundrum model  containing
gravity and scalar field does not have any four-dimensional vector
fields. Models in higher dimensions contain more degrees of freedom
and the appearance of localized vector fields is more probable.
Consider as an example a Nielsen-Olesen string in six dimensions. It
is known that this theory can localize gravity as in the thin string
limit \cite{Gherghetta:2000qi} as well as for a thick string
\cite{Giovannini:2001hh}.

The model contains a U(1) gauge field, a complex scalar field and
gravity with $D_2=1$.  We keep $D_1$ arbitrary. The background
metric has the general form (\ref{simpl}) with $dy= a d\theta$,
$0\leq\theta<2\pi$ being an  angular coordinate and $0\leq r
<\infty$  a radial coordinate, $a$ is a parameter of the dimension
of a length. For a gravity localizing solution $A$ and $B$ go like
$-cr$ for $r \to \infty$, where $c$ is some positive constant. The
background vector field $A_\theta(r) $ and scalar field $\Phi(r)$
are non-zero and are given by a standard Nielsen-Olesen
configuration,
\begin{eqnarray}
&& \Phi(r,\theta) =v f(r)e^{in\,\theta},
\nonumber\\
&&A_{\theta}(r)  =\frac{1}{a e}[P(r)-n] ~, 
\label{NO}
\end{eqnarray}
where $v$ is the vev of the scalar field, and the functions $f(r)$
and $P(r)$ satisfy the following boundary conditions:
\begin{eqnarray}
f(0)=0,&\qquad& \lim_{r\rightarrow \infty} f(r)=1,
\nonumber\\
P(0)=n,&\qquad& \lim_{r\rightarrow \infty} P(r)=0.
\label{boundary}
\end{eqnarray}
Their dependence on $r$ can be found by numerical integration of
equations of motion and is presented in \cite{Giovannini:2001hh}.

In the light-cone gauge one gets three four-dimensional vector
fields, namely $h_{ri},~h_{\theta i}$ and $V_i$. One of the vector
degrees of freedom, defined by (\ref{grvect}), is in fact a part of
four-dimensional massive spin two field and thus is not interesting
for us. A combination of $V_i$ and $h_{\theta i}$, to be identified
below will represent a localized spin one zero mode, found recently
by a direct computation in
\cite{Giovannini:2002sb,Giovannini:2002mk}, while a second
combination will be a massive spin one field. In the absence of the
U(1) gauge field the massless zero mode corresponds to a graviphoton
of U(1) isometry group \cite{Neronov:2001br}.

To determine the structure of zero mode it is helpful
\cite{Randjbar-Daemi:1982hi} to consider a combined action of
linearized  U(1) gauge transformation with the gauge function
$\alpha(x^\mu)$ and a general coordinate transformation of the
U(1) isometry group $\delta (a\theta)=\xi^\theta(x^\mu)$ on the
$(\theta\mu)$ components of the metric and $\mu$ component of the
vector field. Using the transformation rules given in equations
(\ref{gc1}) to (\ref{gc3}) we find
\begin{eqnarray}
\delta h_{\theta\mu}=-e^B\partial_\mu \xi^\theta~,\\
\delta V_\mu =-\partial_\mu \xi^\theta A_\theta -
\frac{1}{e}\partial_\mu\alpha~.
\end{eqnarray}
Under this transformation different scalar fields transform as
\begin{eqnarray}
\delta h_{r\theta}=\delta V_r=0~,\\
\delta \phi=(-\xi^\theta\partial_{a\theta} + i\alpha)\Phi~.
\label{combined}
\end{eqnarray}
From eq. (\ref{combined}) one can easily see that the group
U(1)$_{gauge}\times$U(1)$_{isometry}$ is spontaneously broken down
to U(1)$_\gamma$ that corresponds to a gauge transformation with
$\alpha= \frac{n}{a}\xi^\theta$. Under this transformation the
change of the vector field is
\begin{equation}
\delta V_\mu =-\partial_\mu \xi^\theta (A_\theta +\frac{n}{a e})=
-\frac{1}{ae}\partial_\mu \xi^\theta P(r)~,
\end{equation}
while the field $h_{ir}$ is invariant.  These transformation rules
indicate that the field $\gamma_\mu(x^\nu)$ defined by
\begin{equation}
V_\mu= \frac{1}{ae}P(r)\gamma_\mu,~~
h_{\mu\theta}=  e^B \gamma_\mu
\end{equation}
should correspond to a four-dimensional massless vector field.
Indeed, a direct computation gives for the action of $\gamma_i$:
\begin{equation}
S= -\frac{b}{2}\int d^{D_1}x \eta^{\mu\nu}\partial_\mu\gamma_i\partial_\nu\gamma_i
\label{photon}
\end{equation} 
with a finite normalization factor
\begin{equation}
b=2\pi a\int_0^\infty dr e^{(\frac{D_1}{2}-2)A+\frac{1}{2}B} 
\left(\frac{P^2(r)}{a^2e^2}+ \frac{e^B}{\kappa^2}\right)~.
\end{equation}
In deriving of eq. (\ref{photon}) one also need to use
the difference between the  $00$ and $\theta\theta$ components of
the background equation which takes the form
\begin{equation}
 e^B\left( \frac{1}{2} A'' -\frac{1}{2} B'' +\frac{D_1}{4} {A'}^2-
\frac{1}{4}{B'}^2-\frac{1}{4}(D_1-1)A'B'\right)=
\frac{\kappa^2}{2a^2}(\frac{1}{e^2}{P'}^2+2v^2f^2P^2)~.
\end{equation}
We note that all these integrals are finite even for $D_1 = 4$ for
which the warp factor $e^A$ disappears from the exponential of the
integrand defining $b$.  

\subsection{t'Hooft-Polyakov monopole in seven dimensions}
As was shown in \cite{Gherghetta:2000jf} (see also \cite{ewald}),
localization of gravity is possible on a t'Hooft-Polyakov monopole
configuration provided the model parameters are tuned is a
specific way. For this theory $D_1=4,~D_2=2$, and $g_{mn}(y)dy^m
dy^n= \sin^2\theta d\varphi^2 + d\theta^2$, with
$0\leq\theta\leq\pi$ and $0\leq \varphi\leq 2\pi$ being the
angular variables, whereas $0\leq r <\infty$ is a radial
coordinate. Asymptotics of the functions $A$ and $B$ for $r \to
\infty$ are $A \propto -c r$, $B \to const$. We do not attempt to
make an analysis of the vector field zero mode problem here,
because of the complexity of the monopole solution (see below for
the discussion of an abelian monopole). However, even if the zero
modes exist, they are not normalizable because of the behavior of
$B$ at infinity.

\subsection {$SU(2)\times U(1)$ Gauge Fields}
In this subsection we will study the case of an abelian monopole
configuration and show that it allows the existence of vector zero
modes living in the $x^\mu$ subspace with the symmetry group
SU(2)$\times$U(1). These modes are localized on a brane for $D_1 \geq
5$. For the physically interesting case of $D_1=4$ they are not
normalizable. Perhaps, the localization of these modes may be
achieved by the radiative corrections to the action of the gauge
fields as in \cite{Dvali:2000rx}, or by some non-perturbative
mechanism \cite{Dvali:1996xe,Dubovsky:2001pe}.

Let the gauge group $G=U(1)$ and assume that $D_2=2$ and that the $y$
coordinates cover a $S^2$ with the metric
\begin{equation}
ds^2 = a^2 (d\theta^2  + \sin\theta ^2 d\varphi^2)~,
\end{equation}
where $a$ is the radius of the sphere.  We will assume that there is
a single real scalar field in the system which generates a domain
wall and eventually through their Yukawa coupling to the fermions
produce chiral fermions localized to the wall as we have demonstrated
in \cite{Randjbar-Daemi:2000cr}. The SU(2) gauge fields will have
their origin in the isometries of $S^2$. The gauge field
configuration should thus preserve this symmetry. There is a unique
choice with this symmetry, namely, the magnetic monopole
configuration on $S^2$. This configuration will also be responsible
for the chirality of fermionic fields localized to the brane.

The U(1) monopole configuration is given by
\begin{equation}
 A= \frac{n}{2}(cos\theta -1)d\varphi~,
\end{equation}
where $n$ is an integer.

This solution is one of the class of solutions considered in
\cite{Randjbar-Daemi:2000ft}. There it was shown that the bulk
Einstein and gauge field equation in the absence of the scalar field
are solved by
\begin{equation}
A(r) = -cr, \quad \quad\quad  B(r)=const, \quad\quad F_{nk}F_m^{\quad
k}= f^2 g_{mn}, \quad and\quad  F_{rm=0}~,
\end{equation}
where $c>0$ and $f$ are constants.  Without any loss of generality we
can take $B=0$. The strength of the magnetic field which is given by
the constant $f$  ( $f^2 = \frac{n^2}{2})$ and which should be
quantized for topological reasons.

It is easy to verify that under the assumption that the  real scalar
field approaches the minima of its potential for $r\rightarrow
\pm\infty$ the bulk equations will still  be solve with the above
anzatz.  In this section we shall assume that, with the same anzatz
for the gauge field,  there exists a core solution for the scalar
field which joins smoothly the above bulk solution.  Near the core
the functions $A$ and $B$ will of course be more complicated and will
not be given by simple expressions given above. The details of the
scalar potential or the solution for the functions $\Phi$, $A$ or $B$
will turn out to be irrelevant to the existence of the massless
modes.

Assume that the $\Phi$ field configuration creates a $D_1+2$
dimensional defects located at some points along the $r$
direction. The $D_1$ dimensional subspace is conformal to flat
Minkowski subspace, while the other $2$ directions cover a $S^2$
of radius $a$. To study the possibility of localizing gauge fields
on the $\mathbb{R}^{D_1}$ part we need to look at the fluctuations
around the background solution. Our interest in this section is
only on a sub-sector of the vector part comprising $V_i$ and
$h_{mi}$ subject to the condition $\nabla _m h^{mi}= (\partial_m
+\Gamma_{lm}^{l} )h^m _i=0$. This condition makes the only other
vector field in our problem, namely, $h_{ri}$ to decouple from
this sub-sector. To obtain the spectrum in $\mathbb{R}^{D_1}$ we
perform harmonic expansion on $S^2$. The fields $V_i(x,
\theta,\phi, r)$ and $h_{mi}(x, \theta,\phi,r)$ can be expanded on
spherical modes on $S^2$ according to the prescription given in
\cite{Randjbar-Daemi:1982hi,Randjbar-Daemi:1983bw}. Each expansion
mode will carry a pair of $SU(2)$ labels $l$ ( specifying the
$SU(2)$ irreducible representation) and $\lambda$ taking on $2l+1$
values. For the field $V_i$ the values of $l$ start from zero. The
expansion is given by,
\begin{equation}
  V_i= \sum_{\geq 0}\sqrt{\frac{2l+1}{4\pi}}\sum_{\mid
  \lambda\mid\leq l} V^{l\lambda}_i (x, r)
  D_0 ^{l,\lambda}(\theta,\varphi)~.
\end{equation}
The $l=0$ mode is a $SU(2)$ singlet and it decouples from the rest.
After integrating over $\theta$ and $\phi$ its action becomes
\begin{eqnarray}
S^{l=0} =  a^2\int d^{D_1}x \int dr
e^{(\frac{D_1}{2}-1)A+B}[-\frac{1}{2e^2}e^{-A }\eta^{\mu\nu}
\partial_\mu V_i(x,r)\partial_\nu V_i(x,r) &+&\\
\frac{1}{2e^2} V_i (\partial_r ^{2} +((\frac{D_1}{2} -1)A'+
B')\partial_r )V_i]~.
\end{eqnarray}
The operator $e^{(\frac{D_1}{2}-1)A+B}\left(\partial_r ^{2}
+\left((\frac{D_1}{2} -1)A'+ B'\right)\partial_r\right)$ has the
general structure  of \footnote{ A similar operator exists also
for the string solution discussed in section (6.1).}
$e^N(\partial_r ^{2} + N'\partial_r)$ with $N=
{(\frac{D_1}{2}-1)A+B}$. This operator is a negative Hermitian
operator.  It has two zero modes. The first one is the trivial
one, namely, $\partial_r V_i=0$ This mode obviously produces a
massless gauge field in $\mathbb{R}^{D_1}$ with an effective gauge
coupling constant given by
\begin{equation}
\frac{1}{g^2}=\frac{a^2}{e^2}\int dr e^{(\frac{D_1}{2}-2)A+B}~.
\end{equation}
Provided that this gauge field has the correct coupling to the chiral
fermions localized on $\mathbb{R}^{D_1}$ we can identify it with the
$U(1)_Y$ gauge field of the standard $SU(2)_L\times U(1)_Y$
electroweak theory.

The second zero mode has a non trivial dependence on $r$ and is
given by $V_i (x, y, r)= \int^r dr' e^{ - N(r')}\tilde V_i (x, y)$.
Using this mode the above coupling should be replaced by
\begin{equation}
\frac{1}{g^2}=\frac{a^2}{e^2}\int dr e^{(\frac{D_1}{2}-2)A+B}\psi^2~,
\end{equation}
where $\psi(r) = \int^r dr'e^{- N(r')}$.

The value of the kinetic term for the massless vector or the
effective gauge coupling after $r$ integration will depend on the
details of the background solution. However, there are some general
statements which we can make about the convergence of the integral,
which will also be valid for the $SU(2)$ coupling to be discussed
below. As we are assuming that the defect is created by a scalar
field configuration of a kink type we shall assume that the
coordinate $r$ ranges from $-\infty$ to $+\infty$. The kink will be
assumed to be localized at $r=0$. Away from $r=0$ the scalar field
will approach one of the minima of its potential $U(\Phi)$, and it
will be very slowly varying so that in the background field equations
we can set $\Phi'=0$. As stated above it is easily verified that in
this bulk region the background equations admit a solution of the
form $B=0$ and $A= cr$, where the constant $c$ is related to the
radius of $S^2$ and the magnetic field $f$ through
\begin{equation}
 c^2 = \frac{4}{D_1} \left(\frac{\kappa^2 f^2}{2e^2}-
\frac{1}{a^2}\right)~.
\end{equation}
We thus have two solutions for $c$ differing in sign. Both signs lead
to regular bulk geometries. In the vicinity of the brane the shape of
the functions $A(r)$ and $B(r)$ will depend on the details of the
scalar potential and the detailed form of the scalar field
configuration in that region. Unfortunately, neither the trivial
$r$-zero mode nor the non trivial one given by $\int^r dr' e^{ -
N(r')}$ produce a convergent integral for the gauge coupling for
$D_1=4$. On the other hand the trivial zero mode in the gravitational
sector produces a finite $D_1=4$ effective Newton's
constant\footnote{We can perform similar analysis for the
localization of graviton. The resulting effective $D_1$ dimensional
gravitational constant will be give by $\frac{1}{\kappa_{D_1} ^{2}} =
\frac{a^2}{\kappa^2}\int dr e^{( \frac{D_1}{2}-1)A +B} \Psi^2(r)$
where this time $\Psi$ is the gravitational $r$ zero mode which is
$1$ in the trivial sector and $\Psi (r)= \int^r dr' exp \left( -
\frac{D_1}{2}A-B\right)$ in the non-trivial sector. It is easy to see
that the trivial $r$ zero mode will give a convergent effective $D_1$
dimensional gravitational coupling constant, even for $D_1=4$
provided we select the exponentially decaying asymptotic solution for
the warp factor.}, provided we select the exponentially decreasing
warp factor, i.e in the asymptotic region we set $c= -|c|\varepsilon
(r)$. This feature will persist also for the non-Abelian effective
gauge couplings. We shall offer some speculative ideas which may help
to overcome this important difficulty.

Now consider the $l=1$ sector. This comprises the $SU(2)$ triplets
$h_{i\lambda}$ and $V_{i\lambda}$ contained in the harmonic
expansion of $h_{mi}(x, \theta,\phi, r)$ and $V_i(x, \theta,\phi,
r)$.

According to \cite{Randjbar-Daemi:1982hi} in order to expand
$h_{im}$ into harmonic modes on $S^2$ we need to use  the isohelicity
$\pm$ basis. This is constructed from the components of a vector
relative to some orthonormal frame at any point in the tangent space
to $S^2$. Let $h_{1i},h_{2i}$ denote the components of $h_{mi}$
relative to some orthonormal frame. The complex fields $h_{\pm
i}$ are defined by
\begin{equation}
 h_{\pm i}= \frac{1}{\sqrt{2}}(h_{1i} \mp h_{2i})~.
\end{equation}
Using the formalism of
\cite{Randjbar-Daemi:1982hi,Randjbar-Daemi:1983bw}
we can now write down the following mode
expansion for the fields $h_{\pm i}$
\begin{equation}
  h_{i+} = \bar h_{-i} =\sum_{l\geq
1}\sqrt{\frac{2l+1}{4\pi}}\sum_{\mid
  \lambda\mid\leq l} h_{i+} ^{l\lambda} (x, r)
  D_1 ^{l,\lambda}(\theta,\varphi)~.
\end{equation}

It is not so hard to obtain the bilinear action for the infinite
tower of the Kaluza Klein modes. We give the general result in
Appendix B.  Here we shall restrict our attention to the $l=1$
sector only. The corresponding action in terms of $h_{i\lambda}$ and
$V_{i\lambda}$, where $\lambda$ is the $SU(2)$ triplet index,   is
rather complicated. However, if we  perform  a change of variables
from $h_{i\lambda}$ to $U_{i\lambda} = e^{-B} h_{i\lambda}$ and make
use of the background equations in an efficient way then we obtain
considerable simplification. Again we omit the tedious calculations.
The result is
\begin{eqnarray*}
S_{l=1}(U,V)&= &a^2\int d^{D_1}x\int dr
 e^{(\frac{D_1}{2}-1)A+B} [
 -\frac{e^{-A}}{2e^2}\eta^{\mu\nu}\partial_\mu V_i\partial_\nu V_i
 +\\
 &&+\frac{1}{2e^2}\bar V_i\{\partial^2_{r} +
 ((\frac{D_1}{2}-1)A' +B')\partial_r -
 \frac{2}{a^2}e^{-B}\}V_i +\\
 &&+\frac{\kappa}{e^2}\frac{f}{a}e^{-B}(\bar V_iU_i + V_i\bar U_i)
 -\\
 &&-e^{-A+B} \eta^{\mu\nu}\partial_\mu{\bar U}_i\partial_\nu U_i
 + e^B {\bar U}_i (\partial_{r}^2 + ((\frac{D_1}{2}-1)A'
 +2B')\partial_r
 -\frac{\kappa^2}{e^2}f^2 e^{-2B})U_i]~.
\end{eqnarray*}
Note that we have suppressed the SU(2) triplet index $\lambda$
which is summed over the values $\pm1, 0$. The $r$ derivative
operator acting on $V_i$, $U_i$ have the general structure of $e^N
(\partial_r ^2 +N'\partial_r)$. This is an Hermitian and negative
operator, as we mentioned above. It has again two zero modes, one
with trivial dependence on $r$ and the other with a non trivial $r$
dependence of the type we discussed above. In both of the  zero mode
sectors  the action can be diagonalized. For example in the trivial
sector the simple transformation $V_i= \alpha V'_{i} + \beta U'_i$
and $U_i = \frac{1}{2a\kappa f}\alpha V'{_i} -
\frac{1}{2}\frac{a\kappa f}{e^2}\beta U'{_i}$ where $\alpha$ and
$\beta$ are arbitrary numbers diagonalizes the action. The diagonal
action for both the trivial and non-trivial zero mode sectors take
the form of
\begin{equation}
S_{l=1}(V', U') =\int d^{D_1}x \left(-\frac{1}{2g_{2}
^2}\eta^{\mu\nu} \partial_\mu\bar V_i'\partial_\nu V'_i + \bar U'{_i}
( \frac{c}{2}\eta^{\mu\nu} \partial_\mu\partial_\nu -
m^2)U'_i\right)~.
\end{equation}
Thus the field $V'_{i\lambda}(x_)$ is  a triplet of massless
vectors to be identified with the gauge fields of $SU(2)_L$ with a
coupling constant $g_2$ given by
\begin{equation}
\frac{1}{g_2^{2}} = |\alpha|^2\frac{a^2}{e^2}\int dr e^{(
\frac{D_1}{2}-2)A +B} \Psi^2(r) \left( 1 + \frac{2e^2}{(a\kappa
f)^2} e^{B}\right)~,
\end{equation}
where $\Psi(r) =1$ for the trivial $r$-zero mode and $\Psi (r)=
\int^r dr' exp \left( -( \frac{D_1}{2}-1)A-B\right)$ for the
non-trivial zero mode. It is seen that  neither the trivial nor the
non trivial $r$ zero modes  will cut the $r$ integral when $|r|$ goes
to infinity, for the case of $D_1=4$. The trivial $r$ zero mode
together with asymptotically decaying warp factor, on the other hand
produces convergent integrals for both the Abelian as well as the non
Abelian gauge couplings for any $D_1> 4$ and a finite effective
gravitational constant for any $D_1\geq 4$.

The other constants in $S$ above are defined through,
\begin{equation}
c=\frac{a^2|\beta|^2}{e^2}\int dr e^{( \frac{D_1}{2}-2)A +B}
\Psi^2(r)\left( 1 + \frac{(a\kappa f)^2}{2e^2} e^{-B}\right)~,
\end{equation}
\begin{equation}
\quad\quad\quad\quad\quad  m^2 = \frac{|\beta|^2}{e^2}
\int dr e^{(
\frac{D_1}{2}-1)A }  \Psi^2(r)\left( 1 + \frac{(a\kappa
f)^2}{2e^2} e^{-B} + \frac{(a\kappa f)^4}{4e^4} e^{-2B}\right)~.
\end{equation}
Note that the complex numbers $\alpha$ and $\beta$ are not fixed.
The normalization of the non linear terms in the Yang-Mills action
and the fermion SU(2) coupling constant should fix them.

Thus, together with the singlet massless vector we obtain the four
massless vectors of the standard S(2)$\times$U(1) gauge theory.
We also obtain a SU(2) triplet of massive vectors $U'_i$. It is
hoped that this vector and other massive bulk modes can be made
 arbitrarily heavy or weakly coupled to the standard model
sector.

\subsection{SU(3)$\times$SU(2)$\times$U(1) Gauge Fields}
To obtain color SU(3) gauge fields using the above Kaluza-Klein
mechanism in the context of the brane world scenarios, we add an
extra component to the compact internal space, namely, we take the
$y$ subspace to be  $S^2\times CP^2$. Thus $D_2=6$. It is sufficient
to start from a D-dimensional U(1) gauge theory with a single real
scalar  field which will produce a defect as in the monopole case.
The Einstein, Yang-Mills system will be solved again with the $S^2$
metric as given in the previous section and with the standard
Fubini-Study metric on $CP^2$. Using a pair of complex coordinates
$\zeta^a, a=1,2$ on $CP^2$ this metric is given by
\begin{equation}
 ds^2= \frac{4b^2}{1+\zeta^{\dag}\zeta}
d{\bar \zeta}^a \left(\delta^{ab}-
\frac{\zeta^a{\bar\zeta}^b}{1+\zeta^{\dag} \zeta}
\right)d\zeta^b,
\end{equation}
where $b$ is the radius of $CP^2$. The anzatz for the U(1) gauge
field is a magnetic monopole on $S^2$ and a self -dual instanton on
$CP^2$, which is given by,
\begin{equation}
 A= iq \frac{1}{2\left(1+\zeta^{\dag}
\zeta\right)}\left(\zeta^\dag d\zeta - d\zeta^\dag \zeta
\right)~,
\label{a}
\end{equation}
where $q$ is a real number. The field strength of this vector
potential is self-dual. For this reason we call it a U(1) instanton.
It is well known that in the absence of a U(1) gauge field of the
above type there is a global obstruction to defining a spinor field
on $CP^2$. With the background gauge field as given above this
obstruction is removed provided $q$ is taken to be one half of an odd
integer\footnote{This background is one of the configurations studied
in detail in reference \cite{Dvali:2001qr} in the context of a  new
proposal to solve the gauge hierarchy problem in theories with large
extra dimensions.}.

The configuration of the gauge and the metric fields given above
solve the coupled bulk Einstein, Yang-Mills equations in the region
where the scalar field sits at the minima of its potential.  We
assume that there is a solution in the core of the defect which
smoothly joins the bulk solution as in the monopole case. The
isometry group of the Fubini-Study metric is SU(3). Using the
techniques similar to those outlined for the monopole case above we
will obtain SU(3)$\times$ SU(2)$\times$ U(1) gauge fields in the
subspace spanned by $x^\mu$.

In ref \cite{Dvali:2001qr} it has been shown that the $S^2\times
CP^2$ together with the magnetic monopole on $S^2$ and the self dual
instanton on $CP^2$ gives rise to a rich spectrum of chiral fermions
in the $x$-subspace. One can choose the magnetic charge and the
instanton number such that the chiral fermions belong to the doublet
representation of SU(2) and be at the same time triplets of SU(3).

As in the monopole case in the previous section, these zero modes are
localized on the defect for $D_1 \geq 5$ and not normalizable for
$D_1=4$.

If we wish to obtain all the gauge fields of the standard model from
a geometrical origin the minimum number of dimensions needed in our
scheme turns out to be $4+2+4+1=11$.  At the end of the next section
we speculate on a possible origin for these gauge fields in the
11-dimensional $M$-theory.

\section{Conclusions}

In this paper we gave a detailed analysis of the bosonic fluctuations
of field-theoretical branes -- topological defects. The general
equations we obtained can be used for constructing a low-energy
effective theory coming from the fluctuations around classical
solutions residing in higher-dimensional space-time.

Whereas localization of gravity is possible provided certain
fine-tuning conditions are satisfied, existence of non Abelian
normalizable zero modes for the gauge fields is a serious problem
in the case of $D_1\leq 4$. This is not so for the $U(1)$ gauge
fields as shown in section 6.1. In this section we demonstrated
that in the background of a Nielsen - Olesen string in any $D_1+2$
dimensional theory there is a normalizable $U(1)$ gauge field
localized to the $D_1$ dimensional space-time transverse to the
string for any $D_1$.
 In the monopole background where  non Abelian $D_1$
dimensional gauge fields can be obtained, the spin-1 kinetic
energy of the effective $D_1$ dimensional theory is finite only
for $D_1 > 4$. Perhaps, higher dimensional field theories are not
 relevant for four-dimensional non Abelian physics and a
resolution of the problem of the gauge-field localization should
come from string theory and D-branes.

A way out at the field theory level would be to add, say, to U(1)
Higgs model in six dimensions, forming a string, an extra gauge
group that does not couple to the gauge and Higgs field forming a
topological defect. However, this does not look particularly
natural from model building point of view. Another possibility
would be to modify the action of gravity and/or of the gauge field
by writing, say $F_{AB}^2 Z(\Phi)$ where $Z$ is some function with
the property that it goes to zero when $\Phi$ reaches its
asymptotic value. This does not look very appealing either. So, it
remains to be seen if a natural realistic four-dimensional theory
can be constructed along these lines.

It is possible of course that part or all of the standard model gauge
group emerges as an unbroken subgroup of a big non-abelian gauge
group in higher dimensions. This is what happens in the Calabi-Yau
compactification of gauge theories. Being Ricci flat the Calabi Yau
spaces do not admit continuous isometries. The 4-dimensional gauge
symmetries of particle physics emerge as unbroken subgroups of the
$E_8\times E_8$ group of the heterotic string theory in ten
dimensions. If we insist on a geometrical origin for the gauge
symmetries the minimum number of dimensions in our scheme is 4+
2+4+1=11. This is identical to the number of the dimensions of the
$M$-theory which is believed to have the 11-dimensional supergravity
as its low energy limit. The 11-dimensional supergravity has
neither elementary scalar fields  nor a $U(1)$ gauge field in its
spectrum. Its bosonic spectrum consists of a third rank antisymmetric
potential $A_{MNP}$ and of course the gravity $G_{MN}$. The component
$A_{m\mu\nu}$, where, $m$ is tangent to $S^2$ or to $CP^2$ will look
like a vector field on $S^2$ and a vector field on $CP^2$. It will be
like a scalar field from the point of view of the 4-dimensional space
time covered by the $x^\mu$ coordinates. It can play the role of the
$U(1)$ gauge field in our construction above.  There are also several
candidates for real scalar fields which can depend on the $r$
coordinate only, for example, $A_{\mu\nu\lambda}$ is one of them.
Unfortunately the 11-dimensional supergravity, at least at the
classical level,  does not have  a non zero potential for these
scalars, as the only non linear term involving the tensor field
$A_{MNP}$ is the Chern-Simons term.  It remains to be seen,
nevertheless, if we can generate a solution of the type we need, at
the quantum level.

\noindent {\it Acknowledgments:}
We wish to thank Massimo Giovannini and Peter Tinyakov for helpful
discussions. This work was supported by the FNRS, contract no.
20-64859.01. S.R.D thanks the Institute of Theoretical
Physics in Lausanne and M.S. thanks ICTP for the hospitality.

\newpage
\begin{appendix}

\renewcommand{\theequation}{A.\arabic{equation}}
\setcounter{equation}{0}

\section*{Appendix A: The curvature tensor for a warped metric}

Our convention for the Ricci tensor is $R{\mu\nu}= R^{\lambda}
_{\mu\lambda\nu}$, where the Riemann tensor is defined by
\begin{equation}
R^{\mu}_{\gamma\alpha\beta}= \partial_{\alpha} \Gamma^{\mu}
_{\beta\gamma} -\partial_{\beta}\Gamma^{\mu}_{\alpha\gamma} +
\Gamma^{\mu}_{\alpha\lambda} \Gamma^{\lambda}_{\beta\gamma}-
\Gamma^{\mu} _{\beta\lambda}\Gamma^{\lambda}_{\alpha\gamma}
\end{equation}
For the warped metric (\ref{simpl}) the different non-zero components
of the connections with mixed indicies are:
\begin{equation}
\Gamma_{\mu\nu}^r = -\frac{1}{2}\eta_{\mu\nu}A'e^A~,
\Gamma_{\nu r}^\mu=  \frac{1}{2}\delta^\mu_\nu A'~,
\end{equation}
\begin{equation}
\Gamma_{mn}^r = -\frac{1}{2}g_{mn}B'e^B~,
\Gamma_{n r}^m=  \frac{1}{2}\delta^m_n B'~.
\end{equation}
The curvature and Ricci tensors  are:
\begin{equation}
R_{\mu\nu\rho\sigma}=\frac{1}{4}\left(A'e^A\right)^2
\left(\eta_{\nu\rho}\eta_{\mu\sigma}-
\eta_{\nu\sigma}\eta_{\mu\rho}\right)~,
\end{equation}
\begin{equation}
R_{r\mu r\nu}=-\frac{1}{2} e^A
\eta_{\mu\nu}\left(A'' + \frac{1}{2}{A'}^2\right)~,
\end{equation}
\begin{equation}
R_{m\mu n\nu}=-\frac{1}{4} A' B' e^{A+B}\eta_{\mu\nu} g_{mn}~,
\end{equation}
\begin{equation}
R_{rmrn}=-\frac{1}{2} e^B g_{mn}
\left(B'' + \frac{1}{2}{B'}^2\right)~,
\end{equation}
\begin{equation}
R_{mnkl}=e^B\Omega_{mnkl} +\frac{1}{4}\left(B'e^B\right)^2
\left(g_{nk}g_{ml}-g_{nl}g_{mk}\right)~,
\end{equation}
\begin{equation}
R_{\mu\nu}= -\frac{1}{2}e^{A(r)}\eta_{\mu\nu}\left(A''
+\frac{A'}{2}(D_1 A' + D_2 B')\right)~,
\end{equation}
\begin{equation}
R_{rr}=-\frac{D_1}{2}\left(A''+\frac{1}{2}{A'}^2 \right) -
\frac{D_2}{2}\left(B''+\frac{1}{2}{B'}^2 \right)~,
\end{equation}
\begin{equation}
R_{mn}= \Omega_{mn} - \frac{1}{2}e^{B(r)}g_{mn}\left(B''
+\frac{B'}{2}(D_1 A' + D_2 B')\right)~,
\end{equation}
and the scalar curvature is
\begin{equation}
R=e^{-B(r)}\Omega - D_1 A'' -D_2 B'' -\frac{1}{4}(D_1 A' + D_2 B')^2
-\frac{1}{4}(D_1 {A'}^2 + D_2 {B'}^2)~,
\end{equation}
where $\Omega_{mnkl}$, $\Omega_{mn}$ and $\Omega$ are the curvature,
Ricci and the scalar curvatures constructed from the metric
$g_{mn}(y)$.

\newpage
\renewcommand{\theequation}{B.\arabic{equation}}
\setcounter{equation}{0}

\section*{Appendix B: General Vector Bilinears}

The general vector bilinear in the presence of a neutral scalar can
be shown by some calculation  to be given by (we make judicious use
of the background field equations and rescale $h_{mi}\rightarrow e^B
h_{mi}$ and  assume that $F_{mr}=0$),
\begin{eqnarray*}
\null\hspace{1cm}S = &&\int d^Dx\ e^{\frac{D_1}{2} A + \frac{D_2}{2}
B}  \left[\frac{1}{2e^2}\ e^{-2A}\ V_i \eta^{\mu\nu}\
\partial_\mu\partial_\nu\ V_i\right.
\\[3mm]
&&+\frac{1}{2e^2} \ e^{-A} V_i\left\{ \partial^2_r +
\left(\left(\frac{D_1}{2} -1\right) A'+\frac{D_2}{2}B'\right)\
\partial_r\right. \left. -e^{-B}g^{mn}D_mD_n\right\}\ V_i
\\[3mm]
&&-\frac{\kappa}{e^2} \ e^{-A-B} g^{mn}\ g^{\ell k} D_n V_i \ h_{ki}\
F_{m\ell}
\\[3mm]
&&+\frac{1}{2} \ g^{mn}\ h_{mi}\left\{ \ e^{-2A-B}\ \eta^{\mu\nu}\
\partial_\mu\ \partial_\nu\right.
\\[3mm]
&&+e^{-A+B}\left\{\partial^2_r + \left(\left(\frac{D_1}{2}-1\right)\
A' +\left(\frac{D_2}{2}+1\right)\ B'\right)\ \partial_r\right.
\\[3mm]
&&+e^{-B}\left(\frac{1}{D_2}\Omega +g^{k\ell}\nabla_k\
\nabla_\ell\right) \left. \left.\left. -\frac{\kappa^2 f^2}{e^2}\
e^{-2B}\right.\right\}\ h_{ni}\right]~.
\end{eqnarray*}

In principle we can expand $V_i$ and $h_{mi}$ on the basis of the
eigenfunctions of hermitian differential operators,
$e^{\left(\frac{D_1}{2}-1\right)\ A+\frac{D_2}{2}\
B}\left\{\partial^2_r
+\left(\left(\frac{D_1}{2}-1\right)A'+
\frac{D_2}{2} B'\right)\partial_r\right\}$ acting on $V_i$ and
\begin{equation}
e^{\left(\frac{D_1}{2}-1\right) A+\left(\frac{D_2}{2}+1\right)B}
\left\{ \partial^2_r+\left(\left(\frac{D_1}{2}-1\right)A'+
\left(\frac{D_1}{2}+1\right)B'\right) \partial_r\right\}
\end{equation}
acting on $h_{mi}$.  Both of these operators are of the form
$e^N(\partial^2_r + N'\partial_r)$. Obvious zero modes of these
operators are given by,
\begin{equation}
\partial_r\ V_i = \partial_r\ h_{im}=0~.
\end{equation}
If $r$ ranges from 0, to $\infty$ or $-\infty$  to $\infty$ these
modes are not normalizable in the inner product defined by
$(\psi,\chi) =\int^\infty_0\ dr\ \psi^*\chi$. This may not matter for
the construction of a viable effective theory localized to the
defect.

There is another non trivial  zero mode defined by
\begin{equation}
\left(\partial^2_r + N'\partial_r\right) V_i = 0~,
\end{equation}

\begin{equation}
V_i(x,y,r) = \tilde V_i(x,y)\int^rdr' e^{-N(r')}~.
\end{equation}
Then
\begin{equation}
(V_i,V_i) =\tilde V_i(x,y)\ \tilde V_i(x,y)
\int^\infty dr\ \int^r\ dr'\ e^{-N(r')}
\int^r\ dr'' e^{-N(r'')}~.
\end{equation}
For us $N= \left(\frac{D_1}{2}-1\right) A+\alpha B$, where $\alpha =
\frac{D_2}{2}$ or $\frac{D_2}{2} + 1$.

It is conceivable that there are solutions for which the above
integrals are convergent.

Let us keep the discussion general and write
\begin{equation}
V_i (x,y,r) = e^{\psi(r)}\ \tilde V_i(r,y)~,
\end{equation}

\begin{equation}
h_{im} (x,y,r) = e^{\chi(r)}\ \tilde h _{im} (x,y)~,
\end{equation}
such that  $V_i$ and $h_{im}$ are annihilated by their respective
operators $e^N(\partial^2_r + N'\partial_r)$.

For these zero modes the bilinear action becomes,
\begin{eqnarray*}
S = &&\int d^3 x\ e^{\frac{D_1}{2} A+ \frac{D_2}{2} B}\biggl[\ \biggr.
\tilde V_i\left\{ \frac{e^{-A+2\psi}}{2e^2}  \left(
e^{-A}\eta^{\mu\nu} \partial_\mu \partial_\nu +e^{-B} g^{mn} D_m
D_n\right)\right\}\tilde V_i
\\[3mm]
&&-\frac{\kappa}{e^2} e^{-A-2B+\psi+\chi} g^{mn} g^{k\ell} D_m \tilde
V_i \tilde h_{ni} F_{m\ell}
\\[3mm]
&&+\frac{1}{2} g^{mn} \tilde h_{mi}\left\{
e^{-A+2\chi}\left(e^{-A+B}\eta^{\mu\nu}\partial_\mu\partial_\nu
+\frac{1}{D_2} \Omega +g^{k\ell}\nabla_k\nabla_\ell \left.
-\frac{n^{2}f^2}{e^2}  e^{-B}\right)\right\}\ h_{mi}\right]~.
\end{eqnarray*}
This bilinear action is quite general, applicable to the Abelian and
non-Abelian backgrounds. Now we specialize to the monopole
background.

Take $D_2 = 2$ and $G=U(1)$ with $F_{mn} = f\varepsilon_{mn}$. We
need to make the following substitutions \cite{Randjbar-Daemi:1982hi}:
\begin{equation}
g^{mn}D_mD_n \tilde V_i\to -\frac{1}{a^2} \ell(\ell +1)\tilde V_i\quad
\ell\geq 0~,
\end{equation}
\begin{equation}
g^{mn} D_m D_n \tilde h_{\pm i} = -\frac{1}{a^2} \left\{\ell (\ell
+1)-1\right\}\ \tilde h_{\pm i}\quad \ell \geq 1~.
\end{equation}

Upon integrating over $S^2$ we obtain
\begin{eqnarray*}
S = &&a^2 \int d^{D_1} x \int dr \ e^{\left({\frac{D_1}{2} -
2}\right)A+B+2\psi}\frac{1}{2e^2} \tilde V_i\eta^{\mu\nu}
\partial_\mu\partial_\nu \tilde V_i
\\[3mm]
&&+a^2\int d^{D_1}x \int dr e^{\frac{D_1}{2}A+B} \sum_{\ell\geq
1}\sum_{-\ell\leq \lambda\leq\ell}
\\[3mm]
&&
\overline V_i^{\ell,\lambda} \left\{   \frac{e^{-A+2\psi}}{2e^2}
\left(e^{-A}\eta^{\mu\nu}\partial_\mu\partial_\nu -\frac{1}{a^2}
e^{-B}\ell(\ell+1)\right) V^{\ell,\lambda}_i\right.
\\[3mm]
&&+\frac{\kappa f}{e^2a} \sqrt{\frac{\ell(\ell+1)}{2} }\
e^{-A-B+\psi+\chi}\left(\overline V_i^{\ell,\lambda}\
U^{\ell,\lambda}_i +V^{\ell,\lambda}_i \overline
U^{\ell,\lambda}_i\right)
\\[3mm]
&&+\overline U^{\ell,\lambda}_i\left\{ e^{-A+2\chi} \left ( e^{-A+B}
\eta^{A+B}\eta^{\mu\nu}\partial_\mu\partial_\nu +\right.\right.
\\[3mm]
&&\left.\left.\left.\frac{1}{a^2} - \frac{1}{a^2}
\left(\ell(\ell+1)-1\right)- \frac{\kappa^2f^2}{e^2}
e^{-B}\right)\right\} U^{\ell,\lambda}_i\right\}~,
\end{eqnarray*}
where $U^{\ell,\lambda}_{i}$ are the $SU(2)$ modes of the rescaled
$h_{mi}$. The first term is the $l=0$  contribution given in section
6.3. It is easy to check that the $l=1$ term  also coincides with the
one given in section 6.3.

Although the general result has been obtained with no assumption
about the behavior of the solution near the core of the defect or
defects, it in fact coincides with the result which we would have
obtained in the absence of the scalar field and in the bulk
region. In this case of course the function $B$ has to be set
equal to a constant. In other words the effect of the scalar field
which generates the defects is felt  in the general bilinear
equation above  only through its effect on the functions $A, B$
and the zero mode functions $\psi$ and $\chi$.

\end{appendix}


\newpage

\begin{thebibliography}{99}

\bibitem{kk} Modern Kaluza-Klein Theories (Frontiers in Physics, Vol
65), Editors T. Appelquist, A. Chodos, and P. G.O. Freund.

\bibitem{Witten:1983ux}
E.~Witten,
{\it  in *Appelquist, T. (ed.) et al.: Modern Kaluza-Klein Theories,
438-511.}

\bibitem{Randjbar-Daemi:1982hi}
S.~Randjbar-Daemi, A.~Salam and J.~Strathdee,
Nucl.\ Phys.\ B {\bf 214} (1983) 491.

\bibitem{Randjbar-Daemi:1983bw}
S.~Randjbar-Daemi, A.~Salam and J.~Strathdee,
Phys.\ Lett.\ B {\bf 124} (1983) 345
[Erratum-ibid.\ B {\bf 144} (1984) 455].

\bibitem{Randjbar-Daemi:1983qa}
S.~Randjbar-Daemi, A.~Salam and J.~Strathdee,
Phys.\ Lett.\ B {\bf 132} (1983) 56.

\bibitem{Randjbar-Daemi:1983qb}
S.~Randjbar-Daemi, A.~Salam and J.~Strathdee,
Nucl.\ Phys.\ B {\bf 242} (1984) 447.

\bibitem{GSW:1987}
M.B. Green, J.H. Schwarz and E. Witten
"Superstring Theory", Vol. 2, CUP, 1987.

\bibitem{Rubakov:bb}
V.~A.~Rubakov and M.~E.~Shaposhnikov,
Phys.\ Lett.\ B {\bf 125} (1983) 136.

\bibitem{Akama:jy}
K.~Akama,
Lect.\ Notes Phys.\  {\bf 176} (1982) 267
[arXiv:hep-th/0001113].

\bibitem{Arkani-Hamed:1998rs}
N.~Arkani-Hamed, S.~Dimopoulos and G.~R.~Dvali,
Phys.\ Lett.\ B {\bf 429} (1998) 263
[arXiv:hep-ph/9803315],\\
I.~Antoniadis, N.~Arkani-Hamed, S.~Dimopoulos and G.~R.~Dvali,
Phys.\ Lett.\ B {\bf 436} (1998) 257
[arXiv:hep-ph/9804398].

\bibitem{Antoniadis:1990ew}
I.~Antoniadis,
Phys.\ Lett.\ B {\bf 246} (1990) 377.

\bibitem{Polchinski:1995mt}
J.~Polchinski,
Phys.\ Rev.\ Lett.\  {\bf 75} (1995) 4724
[arXiv:hep-th/9510017].

\bibitem{Randall:1999vf}
L.~Randall and R.~Sundrum,
Phys.\ Rev.\ Lett.\  {\bf 83} (1999) 4690
[arXiv:hep-th/9906064].

\bibitem{Rubakov:1983bz}
V.~A.~Rubakov and M.~E.~Shaposhnikov,
Phys.\ Lett.\ B {\bf 125} (1983) 139.

\bibitem{Gherghetta:2000qi}
T.~Gherghetta and M.~E.~Shaposhnikov,
Phys.\ Rev.\ Lett.\  {\bf 85} (2000) 240
[arXiv:hep-th/0004014].

\bibitem{Gherghetta:2000jf}
T.~Gherghetta, E.~Roessl and M.~E.~Shaposhnikov,
Phys.\ Lett.\ B {\bf 491} (2000) 353
[arXiv:hep-th/0006251].

\bibitem{Giovannini:2001hh}
M.~Giovannini, H.~Meyer and M.~E.~Shaposhnikov,
Nucl.\ Phys.\ B {\bf 619} (2001) 615
[arXiv:hep-th/0104118].

\bibitem{Rubakov:2001kp}
V.~A.~Rubakov,
Phys.\ Usp.\  {\bf 44} (2001) 871
[Usp.\ Fiz.\ Nauk {\bf 171} (2001) 913]
[arXiv:hep-ph/0104152].

\bibitem{Dubovsky:2000am}
S.~L.~Dubovsky, V.~A.~Rubakov and P.~G.~Tinyakov,
Phys.\ Rev.\ D {\bf 62} (2000) 105011
[arXiv:hep-th/0006046].

\bibitem{Dvali:1996xe}
G.~R.~Dvali and M.~A.~Shifman,
Phys.\ Lett.\ B {\bf 396} (1997) 64
[Erratum-ibid.\ B {\bf 407} (1997) 452]
[arXiv:hep-th/9612128].

\bibitem{Dvali:1996bg}
G.~R.~Dvali and M.~A.~Shifman,
Nucl.\ Phys.\ B {\bf 504} (1997) 127
[arXiv:hep-th/9611213].

\bibitem{Dvali:2000rx}
G.~R.~Dvali, G.~Gabadadze and M.~A.~Shifman,
Phys.\ Lett.\ B {\bf 497} (2001) 271
[arXiv:hep-th/0010071].

\bibitem{Dubovsky:2001pe}
S.~L.~Dubovsky and V.~A.~Rubakov,
Int.\ J.\ Mod.\ Phys.\ A {\bf 16} (2001) 4331
[arXiv:hep-th/0105243].

\bibitem{Randjbar-Daemi:2000cr}
S.~Randjbar-Daemi and M.~E.~Shaposhnikov,
Phys.\ Lett.\ B {\bf 492} (2000) 361
[arXiv:hep-th/0008079].

\bibitem{Oda:2000zc}
I.~Oda,
Phys.\ Lett.\ B {\bf 496} (2000) 113
[arXiv:hep-th/0006203].

\bibitem{Bajc:1999mh}
B.~Bajc and G.~Gabadadze,
Phys.\ Lett.\ B {\bf 474} (2000) 282
[arXiv:hep-th/9912232].

\bibitem{Dubovsky:2000av}
S.~L.~Dubovsky, V.~A.~Rubakov and P.~G.~Tinyakov,
JHEP {\bf 0008} (2000) 041
[arXiv:hep-ph/0007179].

\bibitem{Dubovsky:2002xv}
S.~L.~Dubovsky and V.~A.~Rubakov,
arXiv:hep-th/0204205.

\bibitem{Myung:2000dt}
Y.~S.~Myung,
arXiv:hep-th/0009117.

\bibitem{Kehagias:2000au}
A.~Kehagias and K.~Tamvakis,
Phys.\ Lett.\ B {\bf 504} (2001) 38
[arXiv:hep-th/0010112].

\bibitem{Neronov:2001br}
A.~Neronov,
Phys.\ Rev.\ D {\bf 64} (2001) 044018
[arXiv:hep-th/0102210].

\bibitem{Giovannini:2001xg}
M.~Giovannini,
Phys.\ Rev.\ D {\bf 65} (2002) 064008
[arXiv:hep-th/0106131].

\bibitem{Giovannini:2001vt}
M.~Giovannini,
arXiv:hep-th/0111218.

\bibitem{Giovannini:2002jf}
M.~Giovannini,
arXiv:hep-th/0204235.


\bibitem{Giovannini:2001fh}
M.~Giovannini,
Phys.\ Rev.\ D {\bf 64} (2001) 064023
[arXiv:hep-th/0106041].

\bibitem{Randjbar-Daemi:1985wg}
S.~Randjbar-Daemi and C.~Wetterich,
Phys.\ Lett.\ B {\bf 166} (1986) 65.

\bibitem{DeWolfe:1999cp}
O.~DeWolfe, D.~Z.~Freedman, S.~S.~Gubser and A.~Karch,
Phys.\ Rev.\ D {\bf 62} (2000) 046008
[arXiv:hep-th/9909134].

\bibitem{Csaki:2000fc}
C.~Csaki, J.~Erlich, T.~J.~Hollowood and Y.~Shirman,
Nucl.\ Phys.\ B {\bf 581} (2000) 309
[arXiv:hep-th/0001033].

\bibitem{Cohen:1999ia}
A.~G.~Cohen and D.~B.~Kaplan,
Phys.\ Lett.\ B {\bf 470} (1999) 52
[arXiv:hep-th/9910132].

\bibitem{Gregory:1999gv}
R.~Gregory,
Phys.\ Rev.\ Lett.\  {\bf 84} (2000) 2564
[arXiv:hep-th/9911015].

\bibitem{Olasagasti:2000gx}
I.~Olasagasti and A.~Vilenkin,
Phys.\ Rev.\ D {\bf 62} (2000) 044014
[arXiv:hep-th/0003300].

\bibitem{Randjbar-Daemi:1984ap}
S.~Randjbar-Daemi, A.~Salam and J.~Strathdee,
Nuovo Cim.\ B {\bf 84} (1984) 167.

\bibitem{Randjbar-Daemi:1984fs}
S.~Randjbar-Daemi and M.~H.~Sarmadi,
Phys.\ Lett.\ B {\bf 151} (1985) 343.

\bibitem{Lavrelashvili:aa}
G.~V.~Lavrelashvili and P.~G.~Tinyakov,
Sov.\ J.\ Nucl.\ Phys.\  {\bf 41} (1985) 172
[Yad.\ Fiz.\  {\bf 41} (1985) 271].

\bibitem{Randjbar-Daemi:2000ft}
S.~Randjbar-Daemi and M.~E.~Shaposhnikov,
Phys.\ Lett.\ B {\bf 491} (2000) 329
[arXiv:hep-th/0008087].

\bibitem{Cooper:1994eh}
F.~Cooper, A.~Khare and U.~Sukhatme,
Phys.\ Rept.\  {\bf 251} (1995) 267
[arXiv:hep-th/9405029].

\bibitem{Giovannini:2002sb}
M.~Giovannini,
arXiv:hep-th/0205139.

\bibitem{Giovannini:2002mk}
M.~Giovannini, J.~V.~Le Be and S.~Riederer,
arXiv:hep-th/0205222.

\bibitem{ewald}
E.~Roessl and M.~Shaposhnikov,
arXiv:hep-th/0205320.

\bibitem{Dvali:2001qr}
G.~R.~Dvali, S.~Randjbar-Daemi and R.~Tabbash,
Phys.\ Rev.\ D {\bf 65} (2002) 064021
[arXiv:hep-ph/0102307].

\end{thebibliography}
\end{document}